\documentclass[sigconf,natbib,screen=true]{acmart}

\PassOptionsToPackage{dvipsnames}{xcolor}

\usepackage{dsfont}
\usepackage{acronym}
\usepackage{amsmath}
\usepackage[noend]{algorithmic}
\usepackage{algorithm}
\usepackage{bbm}
\usepackage{beramono}
\usepackage{bm}
\usepackage{booktabs}
\usepackage[skip=0pt]{caption}
\usepackage{cleveref}
\usepackage[T1]{fontenc}
\usepackage{graphicx}
\usepackage{hyperref}
\usepackage{listings}
\usepackage{multirow}
\usepackage{natbib}
\usepackage{subcaption}
\usepackage{tabularx}
\usepackage[dvipsnames]{xcolor}
\usepackage{enumerate}
\usepackage[inline]{enumitem}
\usepackage{numprint}
\usepackage[normalem]{ulem}
\usepackage{mleftright}
\usepackage{balance}
\usepackage[subtle]{savetrees}

\setcopyright{rightsretained}

\copyrightyear{2021}
\acmYear{2021}
\setcopyright{rightsretained}
\acmConference[SIGIR '21]{Proceedings of the 44th International ACM SIGIR Conference on Research and Development in Information Retrieval}{July 11--15, 2021}{Virtual Event, Canada}
\acmBooktitle{Proceedings of the 44th International ACM SIGIR Conference on Research and Development in Information Retrieval (SIGIR '21), July 11--15, 2021, Virtual Event, Canada}\acmDOI{10.1145/3404835.3462830}
\acmISBN{978-1-4503-8037-9/21/07}

\ccsdesc[500]{Information systems~Learning to rank}

\keywords{Learning to Rank; Policy Gradients; Ranking Metric Optimization}

\definecolor{rj}{RGB}{0, 150, 0}
\definecolor{mdr}{RGB}{200, 0, 0}
\definecolor{ho}{RGB}{0, 50, 150}

\acrodef{IR}{Information Retrieval}
\acrodef{LTR}{Learning to Rank}
\acrodef{ARP}{Average Relevance Position}
\acrodef{DCG}{Discounted Cumulative Gain}
\acrodef{EM}{Expectation Maximization}
\acrodef{CTR}{Click-Through-Rate}
\acrodef{IPS}{Inverse-Propensity-Scoring}
\acrodef{LogOpt}{Logging-Policy Optimization Algorithm}
\acrodef{RCTR}{Relevant-Click-Through-Rate}
\acrodef{DBGD}{Dueling Bandit Gradient Descent}
\acrodef{COLTR}{Counterfactual Online Learning to Rank}
\acrodef{PDGD}{Pairwise Differentiable Gradient Descent}
\acrodef{NRCTR}{Normalized RCTR}
\acrodef{NDCG}{Normalized DCG}

\acrodef{PL}{Plackett-Luce}

\allowdisplaybreaks

\author{Harrie Oosterhuis}
\affiliation{%
	\institution{Radboud University}
	\city{Nijmegen}
	\country{The Netherlands}
}
\email{harrie.oosterhuis@ru.nl}

\acmSubmissionID{100}

\title[Computationally Efficient Optimization of Plackett-Luce Ranking Models]{Computationally Efficient Optimization\\ of Plackett-Luce Ranking Models for Relevance and Fairness}

\allowdisplaybreaks

\begin{document}

\begin{abstract}
Recent work has proposed stochastic Plackett-Luce (PL) ranking models as a robust choice for optimizing relevance and fairness metrics.
Unlike their deterministic counterparts that require heuristic optimization algorithms, PL models are fully differentiable.
Theoretically, they can be used to optimize ranking metrics via stochastic gradient descent.
However, in practice, the computation of the gradient is infeasible because it requires one to iterate over all possible permutations of items.
Consequently, actual applications rely on approximating the gradient via sampling techniques.

In this paper, we introduce a novel algorithm: PL-Rank, that estimates the gradient of a PL ranking model w.r.t.\ both relevance and fairness metrics.
Unlike existing approaches that are based on policy gradients, PL-Rank makes use of the specific structure of PL models and ranking metrics.
Our experimental analysis shows that PL-Rank has a greater sample-efficiency and is computationally less costly than existing policy gradients, resulting in faster convergence at higher performance.
PL-Rank further enables the industry to apply PL models for more relevant and fairer real-world ranking systems.
\end{abstract}

\fancyhead{}

\maketitle

\acresetall

\section{Introduction}
\label{sec:intro}

\ac{LTR} is a branch of machine learning that covers methods for optimizing ranking systems~\citep{liu2009learning}.
As a result, \ac{LTR} is very important for search and recommendation applications that heavily depend on well-functioning ranking systems.
These systems go through large collections of items and produce a ranking: a small ordered set of items.
A good ranking can make it very easy for the user to find the items they are looking for, even when these items are part of a very large collection~\citep{Chapelle2011, qin2013introducing, dato2016fast}.

Traditionally, ranking systems consist of a scoring function that assigns an individual score to each item, and subsequently, produce rankings by sorting items according to their assigned scores~\citep{liu2009learning, burges2010ranknet, wang2018lambdaloss, joachims2002optimizing}.
The crucial difference with \ac{LTR} and regression or classification is that only the relative differences between scores matter.
In other words, in \ac{LTR} it is not important what the exact score of an item is but how much greater or smaller it is than the scores of the other items.
The main difficulty in \ac{LTR} is that the ranking procedure is deterministic and non-differentiable, since there is no gradient w.r.t.\ the sorting function.
The methods in the \ac{LTR} field can be divided in applying one of two solutions: optimizing a heuristic function that bounds or approximates the ranking performance~\citep{joachims2002optimizing, burges2005learning, liu2009learning, burges2010ranknet, wang2018lambdaloss, bruch2019revisiting}; or optimizing a probabilistic ranking system~\citep{cao2007learning, xia2008listwise, taylor2008softrank, oosterhuis2018differentiable}.

In recent years, the popularity of the \ac{PL} ranking model has increased~\citep{plackett1975analysis, luce2012individual, burges2010ranknet, singh2019policy, diazevaluatingstochastic}.
It models ranking as a succession of decision problems where each individual decision is made by a \ac{PL} model (also known as the \emph{Soft-Max} in deep learning).
Previous research from the industry indicates that the probabilistic nature of the \ac{PL} model leads to more robust performance~\citep{bruch2020stochastic}.
In online \ac{LTR}, it appears the PL model is very good at exploration because it explicitly quantifies its uncertainty~\citep{oosterhuis2018differentiable, oosterhuis2021onlinecounterltr}.
Recent work has also posed that the \ac{PL} model is well suited to address fairness aspects of ranking~\citep{singh2019policy, diazevaluatingstochastic}, because unlike deterministic models, it can give multiple items an equal probability of being the top-item.

However, calculating the gradient of a \ac{PL} ranking model requires an iteration over every possible ranking that the model could produce, i.e., every possible permutation.
In practice this computational infeasibility is circumvented by estimating the gradient based on rankings sampled from the model~\citep{oosterhuis2020taking, oosterhuis2021onlinecounterltr, singh2019policy, diazevaluatingstochastic}.
The main downside of this approach is that it can be computationally very costly.
This is a particular problem in online settings where optimization is performed periodically as more data is gathered~\citep{oosterhuis2020taking, oosterhuis2021onlinecounterltr, singh2019policy, morik2020}.

In this paper, we introduce PL-Rank a novel method that can efficiently optimize both relevance and exposure-based fairness ranking metrics or linear combinations of them.
We contribute to the theory of the \ac{LTR} field, by deriving novel estimators that can unbiasedly estimate the gradient of a \ac{PL} ranking model w.r.t.\ a ranking metric, on which PL-Rank is build.
To the best of our knowledge, PL-Rank is the first \ac{LTR} method that utilizes specific properties of ranking metrics and the PL-ranking model.
Our experimental results show that compared to existing \ac{LTR} methods, PL-Rank has increased sample-efficiency: it requires less sampled rankings to reach the same performance, and increased computational time-efficiency: PL-Rank requires less time to compute the estimation of the gradient and less computational time to converge at optimal performance.
The introduction of PL-Rank makes the optimization of \ac{PL} ranking models more practical by greatly reducing its computational costs, additionally, these gains also help in the further promotion of fairness aspects of ranking models~\citep{diazevaluatingstochastic}.

\section{Related Work}
\label{sec:relatedwork}

One of the earliest \ac{LTR} approaches is the pairwise approach where the loss function is based on the order of pairs of items~\citep{joachims2002optimizing, burges2005learning, liu2009learning}.
While pairwise losses are easy to compute and scale well with the number of items to rank, pairwise loss functions do not consider the entire ranking~\citep{burges2010ranknet, liu2009learning}.
As a result, minimizing a pairwise loss often does not translate to the optimal ranking behavior.
Subsequently, the idea of a listwise method that considers the complete ranking was introduced with the ListNet and ListMLE methods~\citep{cao2007learning, xia2008listwise}.
These methods optimize \ac{PL} ranking models to maximize the probability of the optimal ranking.
They have three main limitations:
\begin{enumerate*}[label=(\roman*)]
\item They assume there is a single optimal ranking per query, while often there are multiple optimal rankings.
\item They are not based on actual ranking metrics and thus may not actually maximize the desired metrics over the entire dataset.
\item They bring substantial computational costs, i.e.\ the cost of ListNet is so high \citeauthor{cao2007learning} only optimize the top-1 ranking~\citep{cao2007learning}.
\end{enumerate*}
Some of these issues are avoided by the later \emph{LambdaRank} and \emph{LambdaMART} methods~\citep{burges2010ranknet}.
LambdaRank is an extension of the pairwise \emph{RankNet} method, where the loss function weights each pair by the absolute difference in \ac{DCG} that would result from swapping the pair.
This approach optimizes a deterministic ranking model and also works for other ranking metrics than \ac{DCG}~\citep{burges2010ranknet, wang2018lambdaloss}.
The Lambda methods are listwise because their gradient is based on the ranking metric values of the current ranking and therefore consider the complete ranking.
While initially, there was only empirical evidence for the great performance of Lambda methods in optimizing ranking metrics~\citep{burges2010ranknet, donmez2009local}.
Recently, \citet{wang2018lambdaloss} proved that the Lamba methods optimize a lower bound on ranking metrics and introduced a novel bound in the form of the LambdaLoss framework~\citep{wang2018lambdaloss}.
Thus although the Lamba methods are based around ranking metrics and are computationally feasible, they do not optimize metrics directly and are therefore heuristic methods.
Another heuristic approach is to replace the rank function in a metric by a differentiable probabilistic approximation, notable examples of this approach are \emph{SoftRank}~\citep{taylor2008softrank} and \emph{ApproxNDCG}~\citep{bruch2019revisiting}.
While all of these methods can be useful in practice, none optimize rankings metrics directly in a computationally feasible manner.

Interestingly, multiple lines of previous work have found \ac{PL}-ranking models to be very effective for various ranking tasks: for result randomization in interleaving~\citep{hofmann2011probabilistic}, multileaving~\citep{schuth2015probabilistic} and counterfactual evaluation~\citep{oosterhuis2020taking}; for exploration in online \ac{LTR}~\citep{oosterhuis2021onlinecounterltr, oosterhuis2018differentiable}; for fair distributions of attention exposure~\citep{singh2019policy, diazevaluatingstochastic}; and for topic diversity in ranking~\citep{xia2017adapting, wei2017reinforcement}.
In particular, \citet{bruch2020stochastic} argue that the stochastic nature of \ac{PL} models results in more robust ranking performance.
Furthermore, \citeauthor{bruch2020stochastic} show that, with small alterations, many existing \ac{LTR} methods can adequately optimize \ac{PL} methods.
An interesting property of the \ac{PL} ranking model is that it has a gradient w.r.t.\ ranking metrics but that it is generally infeasible to compute, existing work has thus approximated this gradient~\citep{oosterhuis2020taking, oosterhuis2021onlinecounterltr, singh2019policy, diazevaluatingstochastic}~(see Section~\ref{sec:plmodels}).
The computational costs are particularly relevant because the \ac{PL} model is often used in online settings where it optimization is performed repeatedly and frequently~\citep{oosterhuis2020taking, oosterhuis2021onlinecounterltr, singh2019policy}.
For instance, \citet{oosterhuis2021onlinecounterltr} show that frequently optimizing the logging-policy model during the gathering of data greatly reduces the data-requirements for online/counterfactual \ac{LTR}.
Similarly, \citet{morik2020} show that to prevent very unfair distributions of exposure, rankings should be updated continuously as more interaction data is gathered. 
To the best of our knowledge, no previous work has developed a novel \ac{LTR} method specifically for \ac{PL} ranking models that optimizes ranking metrics directly nor with a focus on computational efficiency.

\section{Relevance Ranking Metrics}
\label{sec:relevance}

Generally, \ac{LTR} methods assume each item $d$ has some \emph{relevance} w.r.t.\ a query $q$~\citep{liu2009learning}, in the context of fairness this is often considered the \emph{merit} of an item~\citep{diazevaluatingstochastic}.
This is often modelled as the probability that a user finds the item $d$ relevant for their issued query $q$.
To keep our notation brief, we use $\rho_d$ to denote this probability:
$
P(R=1 \,|\, q, d) = \rho_{d}.
$
Whenever we talk about the relevance of an item, it will be clear from the context what the corresponding query is, hence we keep $q$ out of our notation for the sake of brevity.

Rankings are ordered lists of items, we use $y$ to denote a ranking and $y_k$ for the $k$th item in ranking $y$:
$
y = [y_1, y_2, \ldots, y_K],
$
thus $y_k = d$ means that $d$ is the item at rank $k$ in ranking $y$.
We will assume that all rankings are of length $K$, for instance, because only $K$ items can be displayed.
A ranking model $\pi$ can be seen as a distribution over rankings, where $\pi(y\mid q)$ indicates the probability that ranking $y$ is sampled for query $q$ by model $\pi$.
For brevity, we will use $\pi(y) = \pi(y\mid q)$ as the corresponding query will always be clear from the context.

The relevance performance (also called the reward) of a ranking model is represented by the metric value $\mathcal{R}$.
Relevance ranking metrics use weights per rank $\theta$ where the relevance of an item at rank $k$ is weighted by $\theta_k$.
For a single query, the metric is computed as an expectation over the ranking behavior of $\pi$:
\begin{equation}
\begin{split}
\mathcal{R}(q)
&= \sum_{y \in \pi} \pi(y \,|\, q) \sum_{k=1}^K \theta_k P(R=1 \,|\, q, y_k) 
\\
&= \sum_{y \in \pi} \pi(y) \sum_{k=1}^K \theta_k \rho_{y_k}
= \mathbb{E}_y\mleft[\sum_{k=1}^K \theta_k \rho_{y_k}\mright]. \label{eq:queryreward}
\end{split}
\end{equation}
By choosing $\theta_k$ accordingly, $\mathcal{R}(q)$ can represent the most common relevance ranking metrics.
For example, top-$K$ \ac{DCG} is computed with the following weights: 
$
\theta_k^{\text{DCG@K}} =
\frac{\mathds{1}[ k \leq K]}{\log_2(k +1)}
\label{eq:dcg@5}
$,
precision at $K$ with:
$
\theta_k^{\text{PREC@K}} = \frac{1}{K} \mathds{1}[ k \leq K]
$,
or \ac{ARP} can be represented by:
$
\theta_k^{\text{ARP}} = -k
$. %
The overall relevance of a ranking system $\pi$ is simply the expected performance over the distribution of user-issued queries:
\begin{equation}
\mathcal{R} = \mathbb{E}_q\mleft[\mathcal{R}(q)\mright] = \sum_{q \in \mathcal{Q}} P(q) \mathcal{R}(q).
\end{equation}
The value of $\mathcal{R}$ is also often called the \emph{performance} or the \emph{reward}.
Accordingly, \ac{LTR} for relevance optimizes $\pi$ to maximize $\mathcal{R}$, given the item relevances $\rho_d$ and according to the chosen metric represented by $\theta_k$.

\section{Plackett-Luce Ranking Models}
\label{sec:plmodels}

As noted in Section~\ref{sec:relatedwork}, the \ac{PL} model~\citep{plackett1975analysis, luce2012individual} has often been deployed to model a probabilistic distribution over rankings~\citep{hofmann2011probabilistic, oosterhuis2020taking, oosterhuis2021onlinecounterltr, oosterhuis2018differentiable, singh2019policy, diazevaluatingstochastic, xia2017adapting, wei2017reinforcement, bruch2020stochastic}.
In the \ac{PL} model, an item is chosen from a pool of available items based on the individual scores each item has.
For our ranking problem, a learned prediction model $m$ predicts the log score of an item $d$ w.r.t. to query $q$ as $m(q,d) \in \mathds{R}$.
For brevity, we again keep the query out of our notation: $m(q,d) = m(d)$.

The probability that item $d$ is chosen to be the $k$th item in ranking $y$ from the set of items $\mathcal{D}$ is the score of $d$: $e^{m(d)}$, divided by the sum of scores for the items that have not been placed yet:
\begin{equation}
 \pi(d \,|\, y_{1:k-1}, \mathcal{D}) = \frac{e^{m(d)} \mathds{1}[d \not\in y_{1:k-1}]}{\sum_{d' \in D \setminus y_{1:k-1}} e^{m(d')}},
\end{equation}
where $y_{1:k-1}$ indicates the ranking up to rank $k-1$, i.e.,
\begin{equation}
y_{1:k-1} = [y_1, y_2, \ldots, y_{k-1}].
\end{equation}
As such the placement probabilities at $k$ depend only on the scores of the items not placed before rank $k$.
Because the learned $m$ predicts the log score, the actual score is always greater than zero: $e^{m(d)} > 0$, consequently, $\pi(d \,|\, y_{1:k-1})$ is always a valid probability distribution over the unplaced items.
We note that this probability is extremely similar to the Soft-Max function commonly used in deep learning.
To prevent items from appearing in a ranking twice, the probability of $\pi(d \,|\, y_{1:k-1}, \mathcal{D}) = 0$ if $d$ has already been placed: $d \in y_{1:k-1}$.

Finally, the probability of a ranking is simply the product of the placement probabilities of each individual item:
\begin{equation}
\pi(y)  = \prod_{k=1}^K \pi(y_k \,|\, y_{1:k-1}, \mathcal{D}). \label{eq:rankingprod}
\end{equation}

\subsection{Computationally Efficient Sampling}
\label{sec:gumbelsampling}
An advantage of the \ac{PL} ranking model is that rankings can be sampled quite efficiently.
At first glance, sampling a ranking may seem computationally costly, as it involves repeatedly  sampling from the item distribution and renormalizing it.
However, using the Gumbel Softmax trick~\citep{gumbel1954statistical} one can sample an entire ranking without having to calculate any of the actual probabilities~\citep{bruch2020stochastic}.

Our goal is to acquire a sampled ranking $y^{(i)}$ from the $\pi$ distribution:
$y^{(i)} \sim \pi$.
Instead of calculating the actual placement probabilities, for each item a sample from the Gumbel distribution is taken:
$
\gamma^{(i)}_d \sim \text{Gumbel}(0,0).
$
This can be done by first sampling uniformly from the $[0,1]$ range: $\zeta^{(i)}_d \sim \text{Uniform}(0,1)$, and then applying: $\gamma^{(i)}_d = -\log(-\log(\zeta^{(i)}_d))$.
Subsequently, per item we take the sum of their Gumbel sample and their log score:
\begin{equation}
\hat{m}^{(i)}_d = m(d) + \gamma^{(i)}_d.
\end{equation}
Finally, we sort the items according to their $\hat{m}^{(i)}$ values, resulting in the sampled ranking:
\begin{equation}
\begin{split}
y^{(i)}& = [y_1^{(i)}, y_2^{(i)}, \ldots, y_K^{(i)}]
\\
&\text{s.t. } \,\,\, \forall (y_x^{(i)}, y_z^{(i)}), \quad x < z \rightarrow \hat{m}^{(i)}_{y_x} \geq \hat{m}^{(i)}_{y_z}.
\end{split}
\end{equation}
This sampling procedure follows the \ac{PL} distribution of $\pi$~\citep{gumbel1954statistical, bruch2020stochastic}. 
In practice, this means we can sample rankings as quickly as we can sort top-$K$ rankings, which translates to a computational complexity of $\mathcal{O}(|\mathcal{D}| \log( |\mathcal{D}|))$.

\subsection{Basic Policy Gradient Estimation}
\label{sec:basicpolicygradient}
As noted by~\citet{singh2019policy} and~\citet{bruch2020stochastic}, \ac{PL} ranking models can be optimized via policy-gradients.
They utilize the famous \emph{log-trick} from the \emph{REINFORCE} algorithm~\citep{williams1992simple}.
We apply the log-trick to Eq.~\ref{eq:rankingprod} to obtain:
\begin{equation}
\frac{\delta}{\delta m} \pi(y) = 
\pi(y) \mleft[ \frac{\delta}{\delta m} \log(\pi(y)) \mright].
\end{equation}
By combining this result with Eq.~\ref{eq:queryreward}, we find that the derivate can be expressed as an expectation over the ranking distribution $\pi$:
\begin{equation}
\begin{split}
\frac{\delta}{\delta m} \mathcal{R}(q) &= 
\sum_{y \in \pi} \mleft[\frac{\delta}{\delta m} \pi(y)\mright] \sum_{k=1}^K \theta_k \rho_{y_k}
\\
&= \sum_{y \in \pi} \pi(y) 
\mleft[ \frac{\delta}{\delta m} \log(\pi(y)) \mright]
\mleft( \sum_{k=1}^K \theta_k \rho_{y_k} \mright)
\\
&= \mathbb{E}_y \Bigg[ \underbrace{
\vphantom{\mleft(\sum_{x=k}^K \theta_x \rho_{y_x}\mright)}
\mleft[ \frac{\delta}{\delta m} \log(\pi(y)) \mright]
}_{\text{gradient w.r.t.\ complete ranking}}
\underbrace{\mleft( \sum_{k=1}^K \theta_k \rho_{y_k} \mright)}_{\text{full reward}}
\Bigg].
\end{split}
\label{eq:policygradient}
\end{equation}
We see that this policy gradient is composed of two parts: a gradient w.r.t.\ the log probability of a complete ranking multiplied by the reward for that ranking.
In practice, it is infeasible to compute this gradient exactly since it requires a summation over every possible ranking $y$.
Luckily, because the gradient can be expressed as an expectation over ranking $y$ w.r.t.\ to the distribution according to $\pi$, the gradient can be estimated using a simple sampling strategy.
If we sample $N$ rankings from $\pi$ for query $q$, with $y^{(i)}$ denoting the $i$th sample, then the gradient can be estimated using:
\begin{equation}
\begin{split}
\frac{\delta}{\delta m} \mathcal{R}(q) \approx \frac{1}{N}
 \sum_{i=1}^N 
\mleft[ \frac{\delta}{\delta m} \log(\pi(y^{(i)})) \mright]
\mleft( \sum_{k=1}^K \theta_k \rho_{y_k^{(i)}} \mright).
\end{split}
\label{eq:basicapproximation}
\end{equation}
A straightforward implementation first samples $N$ rankings using Gumbel sampling (Section~\ref{sec:gumbelsampling}), and then computes the reward for each ranking, to finally use a machine learning framework to compute the gradient w.r.t.\ the log probabilities: $\mleft[ \frac{\delta}{\delta m} \log(\pi(y^{(i)})) \mright]$.
This approach works well with currently popular deep-learning frameworks such as \emph{PyTorch}~\citep{paszke2019pytorch} or \emph{Tensorflow}~\citep{abadi2016tensorflow}.

This concludes our description of the basic policy gradient approach to optimizing \ac{PL} ranking models.
While this approach works adequately, our results show that this approach is computationally expensive and can have convergency issues when $N < 1000$.
Finally, we note that this approach is not specific to \ac{PL}-ranking models as it essentially just applies the very general \emph{REINFORCE} algorithm~\citep{williams1992simple}.
In contrast, the remainder of this paper will introduce methods that make use of specific \ac{PL} properties, and as a result, show better performance in our experimental results.

\section{Method: PL-Rank for Relevance}
\label{sec:method}

In this section, we will derive three novel methods for estimating the gradient of a \ac{PL}-ranking model.
The latter two estimators make our proposed PL-Rank method, the former is an intermediate step between the basic policy gradient estimation and PL-Rank.
Unlike existing methods, PL-Rank utilizes specific properties about PL ranking models and ranking metrics.

\subsection{Ranking Metric Based Approximation}
\label{sec:placementpolicygradient}

The basic estimator in Eq.~\ref{eq:basicapproximation} only deals with the reward of the entire ranking.
This can lead to very unintuitive behavior, for instance, when a ranking is sampled that receives a very high reward but only due the first placed item, the gradient w.r.t.\ entire ranking will be multiplied with this reward.
Thus despite the fact that only the first item contributed positively to the reward, the probability of placement for all items will be increased.

By rewriting Eq.~\ref{eq:queryreward} we can see that relevance rewards only need to interact with the probability of the ranking up to the corresponding rank:
\begin{equation}
\begin{split}
\mathcal{R}(q)
&= \sum_{y \in \pi} \pi(y) \sum_{k=1}^K \theta_k \rho_{y_k}
=
\sum_{k=1}^K \theta_k \sum_{y \in \pi} \pi(y) \rho_{y_k}
\\
&=
\sum_{k=1}^K \theta_k \sum_{y_{1:k} \in \pi} \pi(y_{1:k}) \rho_{y_k},
\end{split}
\label{eq:rewardpartialprob}
\end{equation}
where $\sum_{y_{1:k} \in \pi}$ is a summation over all possible (sub)rankings of length $k$ according to $\pi$.
In other words, the relevance $\rho_{y_k}$ at any rank $k$ only interacts with the probability of the ranking up to $k$: $\pi(y_{1:k})$.
Inuitively this makes sense because the placement of any item after $k$ will not affect the previously obtained reward.
We can use this fact when estimating the gradient w.r.t.\ the complete reward.

Before we derive the gradient w.r.t.\ the complete reward, we first consider that the derivate of the log probability of a ranking can be decomposed as a sum over log probabilities of the individual item placements.
Using Eq.~\ref{eq:rankingprod}:
\begin{equation}
\begin{split}
\mleft[\frac{\delta}{\delta m}  \pi(y_{1:k}) \mright]
&= \pi(y_{1:k}) \mleft[\frac{\delta}{\delta m}  \log(\pi(y_{1:k})) \mright]
\\
&= \pi(y_{1:k}) \sum_{x=1}^k \mleft[\frac{\delta}{\delta m}  \log(\pi(y_x \mid y_{1:x-1})) \mright].
\end{split}
\label{eq:partialprobderivative}
\end{equation}
We can now use to get the derivative w.r.t.\ to $\mathcal{R}(q)$ using Eq.~\ref{eq:rewardpartialprob}~\&~\ref{eq:partialprobderivative}:
\begin{align}
\nonumber
\frac{\delta}{\delta m} 
\mathcal{R}(q) 
&=
\sum_{k=1}^K \theta_k \sum_{y_{1:k} \in \pi}  \rho_{y_k} \mleft[\frac{\delta}{\delta m}  \pi(y_{1:k}) \mright]
\label{eq:policyplacementderivative} \\ \nonumber
&=
\sum_{k=1}^K \theta_k \sum_{y_{1:k} \in \pi} \pi(y_{1:k})  \rho_{y_k} \sum_{x=1}^k \mleft[\frac{\delta}{\delta m}  \log(\pi(y_x | y_{1:x-1})) \mright]
\\ 
&=
\sum_{k=1}^K  \mathbb{E}_{y_{1:k}} \mleft[ \theta_k
\rho_{y_k} \sum_{x=1}^k \mleft[\frac{\delta}{\delta m}  \log(\pi(y_x \mid y_{1:x-1})) \mright]
\mright]
\\ \nonumber
&=
\mathbb{E}_{y} \mleft[
\sum_{k=1}^K \theta_k
\rho_{y_k} \sum_{x=1}^k \mleft[\frac{\delta}{\delta m}  \log(\pi(y_x \mid y_{1:x-1})) \mright]
\mright]
\\ \nonumber
&=
\mathbb{E}_{y} \Bigg[
\sum_{k=1}^K
\underbrace{
\vphantom{\sum_{x=k}^K \theta_x
\rho_{y_x}} %
\mleft[\frac{\delta}{\delta m}  \log(\pi(y_k \mid y_{1:k-1})) \mright]
}_\text{grad.\  w.r.t.\ item placement }
\underbrace{
\mleft(
\sum_{x=k}^K \theta_x
\rho_{y_x}
\mright)
}_\text{ following reward}
\Bigg],
\end{align}
note that we use following: $\sum_{k=1}^K \mathbb{E}_{y_{1:k}}\mleft[ f(y_{1:k}) \mright] = \mathbb{E}_{y}\mleft[ \sum_{k=1}^K f(y_{1:k}) \mright]$, to move the expectation from partial rankings to complete rankings.
Eq.~\ref{eq:policyplacementderivative} shows us that the derivative consists of two parts: the gradient w.r.t.\ individual item placements and the reward received following each placement.
Again, this gradient can be estimated using rankings sampled from $\pi$:
\begin{equation}
\frac{\delta}{\delta m} 
\mathcal{R}(q)
\approx
\frac{1}{N}
\sum_{i=1}^N
\sum_{k=1}^K
\mleft[\frac{\delta}{\delta m}  \log(\pi(y_k^{(i)} | y_{1:k-1}^{(i)})) \mright]
\sum_{x=k}^K \theta_x
\rho_{y_x^{(i)}}.
\label{eq:placementapproximation}
\end{equation}
We will call this estimator the \emph{placement policy gradient estimator},
in contrast with the basic policy gradient estimator (Eq.~\ref{eq:basicapproximation}), this estimator weights the gradients of placement probabilities with the observed following rewards.
By doing so, it makes use of the structure of ranking metrics and thus is more tailored towards these metrics than the basic estimator.

\subsection{Computationally Efficient Estimation}

So far our placement policy gradient estimator has made use of the fact that the probability of a ranking is a product of individual placement probabilities, however, it has made no further use of the fact that $\pi$ is a \ac{PL} ranking model.
We will now show that using the knowledge that $\pi$ is a \ac{PL} model can lead to an estimator that can be computed with greater computational efficiency.
We start by taking the derivative of an item placement probability:
\begin{align}
\label{eq:placementprobstep}
\frac{\delta}{\delta m} &\pi(d \mid y_{1:k-1})
= \\&
\pi(d \mid y_{1:k-1})
\mleft(
\mleft[\frac{\delta}{\delta m} m(d) \mright] 
- \sum_{d' \in \mathcal{D}} \pi(d' \mid y_{1:k-1}) \mleft[\frac{\delta}{\delta m} m(d') \mright]
\mright).
\nonumber
\end{align}
We note that the probability of placing an item that has already been placed is zero:
$d \in y_{1:k-1} \rightarrow \pi(d \mid y_{1:k-1}) = 0$.
Combining Eq.~\ref{eq:policyplacementderivative}~\&~\ref{eq:placementprobstep} results in the following gradient:
\begin{align}
\frac{\delta}{\delta m} 
\mathcal{R}(q)
&=
\mathbb{E}_{y} \Bigg[
\sum_{k=1}^K
\vphantom{\sum_{x=k}^K \theta_x
\rho_{y_x}} %
\mleft[\frac{\delta}{\delta m}  \log(\pi(y_k \mid y_{1:k-1})) \mright]
\sum_{x=k}^K \theta_x
\rho_{y_x}
\Bigg]
\nonumber \\
&=
\mathbb{E}_{y} \Bigg[
\mleft( \sum_{k=1}^K
\mleft[\frac{\delta}{\delta m} m(y_k) \mright] \mleft( \sum_{x=k}^K \theta_x
\rho_{y_x}
\mright)
\mright)
\label{eq:plrankstep1}
\\ 
&\hspace{0.35cm}
-
\Bigg( \sum_{k=1}^K
\sum_{d' \in \mathcal{D}} \pi(d' \mid y_{1:k-1}) \mleft[\frac{\delta}{\delta m} m(d') \mright]
\Bigg)
\mleft( \sum_{x=k}^K \theta_x
\rho_{y_x}
\mright)
\Bigg].
\nonumber
\end{align}
For the sake of simplicity, we will further derive the resulting two parts of Eq.~\ref{eq:plrankstep1} separately, starting with the first part:
\begin{equation}
\begin{split}
\mathbb{E}_{y} &\Bigg[
\mleft( \sum_{k=1}^K
\mleft[\frac{\delta}{\delta m} m(y_k) \mright] \mleft( \sum_{x=k}^K \theta_x
\rho_{y_x}
\mright)
\mright)
\Bigg]
\\ &=
\mathbb{E}_{y} \Bigg[
\sum_{d \in \mathcal{D}} \mleft[\frac{\delta}{\delta m} m(d) \mright]
\mleft( \sum_{k=1}^K
\mathds{1}[y_k = d]
 \mleft( \sum_{x=k}^K \theta_x
\rho_{y_x}
\mright)
\mright)
\Bigg]
\\ &=
\mathbb{E}_{y} \Bigg[
\sum_{d \in \mathcal{D}} \mleft[\frac{\delta}{\delta m} m(d) \mright]
 \mleft( \sum_{k=1}^K
 \mathds{1}[d \in y_{1:k}]
 \theta_k
\rho_{y_k}
\mright)
\Bigg]
\\ &=
\sum_{d \in \mathcal{D}} \mleft[\frac{\delta}{\delta m} m(d) \mright]
\mathbb{E}_{y} \Bigg[\sum_{k=\text{rank}(d, y)}^K
 \theta_k
\rho_{y_k}
\Bigg].
\end{split}
\label{eq:plrankstep2}
\end{equation}
We see that this first part results in summing over the derivatives of each item score according to model $m(d)$ weighted by the reward expected to follow a placement of $d$.

Then for the second part of Eq.~\ref{eq:plrankstep1}:
\begin{equation}
\begin{split}
&\mathbb{E}_{y} \Bigg[
 \sum_{k=1}^K
\sum_{d \in \mathcal{D}} \pi(d \mid y_{1:k-1}) \mleft[\frac{\delta}{\delta m} m(d) \mright]
\mleft( \sum_{x=k}^K \theta_x
\rho_{y_x}
\mright)
\Bigg]
\\
&\,=
\mathbb{E}_{y}\Bigg[
\sum_{d \in \mathcal{D}} \mleft[\frac{\delta}{\delta m} m(d) \mright]
\sum_{k=1}^K 
 \pi(d \mid y_{1:k-1})
\mleft( \sum_{x=k}^K \theta_x
\rho_{y_x}
\mright)
\Bigg]
\\
&\,=
\sum_{d \in \mathcal{D}} \mleft[\frac{\delta}{\delta m} m(d) \mright]
\mathbb{E}_{y}\Bigg[
\sum_{k=1}^{\text{rank}(d, y)}
 \pi(d \mid y_{1:k-1}) 
\mleft( \sum_{x=k}^K \theta_x
\rho_{y_x}
\mright)
\Bigg],
\end{split}
\label{eq:plrankstep3}
\end{equation}
where we used the fact that:
$k > \text{rank}(d, y) \rightarrow \pi(d \mid y_{1:k-1}) = 0$.
We see that the second part sums over each rank where it multiplies the expected probability that an item was added with the expected following reward.
This product represents the \emph{risk} imposed by an item $d$: if $d$ is not placed at $k$ then $\pi(d \mid y_{1:k-1})$ indicates how likely $d$ would have been placed instead of $y_k$ and in which case the following reward $\sum_{x=k}^K \theta_x\rho_{y_x}$ may not have occurred.
For cases where $d$ is the item at rank $k$: $d = y_k$, the risk stops the log score $m(d)$ from increasing too far as the placement probability $\pi(d \mid y_{1:k-1})$ may already be very great.
By combining Eq.~\ref{eq:plrankstep1},~\ref{eq:plrankstep2}~\&~\ref{eq:plrankstep3} we obtain the full derivative:
\begin{equation}
\begin{split}
\frac{\delta}{\delta m} 
\mathcal{R}(q)
&=
\sum_{d \in \mathcal{D}} \mleft[\frac{\delta}{\delta m} m(d) \mright]
\mathbb{E}_{y} \Bigg[
\overbrace{
\mleft(\sum_{k=\text{rank}(d, y)}^K
 \theta_k
\rho_{y_k}
\mright)
}^\text{reward following placement}
\\
&\qquad\qquad
-
\underbrace{
\sum_{k=1}^{\text{rank}(d, y)}
 \pi(d \mid y_{1:k-1}) 
\mleft( \sum_{x=k}^K \theta_x
\rho_{y_x}
\mright)
}_\text{risk imposed by placement probability}
\Bigg].
\end{split}
\label{eq:plrank1fullderiv}
\end{equation}
We see that the derivative multiplies the gradient of the item log score $m(d)$ with the expected reward following its placement minus the expected risk imposed by $d$ before it is placed.
Finally, this gradient can also be estimated using $N$ sampled rankings:
\begin{equation}
\begin{split}
\frac{\delta}{\delta m} 
\mathcal{R}(q)
&\approx
\frac{1}{N}
\sum_{d \in \mathcal{D}} \mleft[\frac{\delta}{\delta m} m(d) \mright]
\sum_{i=1}^N
\mleft(\sum_{k=\text{rank}(d, y^{(i)})}^K
 \theta_k
\rho_{y_k^{(i)}}
\mright)
\\
&\qquad\quad\,
-
\sum_{k=1}^{\text{rank}(d, y^{(i)})}
 \pi(d \mid y_{1:k-1}^{(i)}) 
\mleft( \sum_{x=k}^K \theta_x
\rho_{y_x^{(i)}}
\mright)
.
\end{split}
\label{eq:plrank1}
\end{equation}
We call this estimator \emph{PL-Rank-1}, to the best of our knowledge this is the first gradient estimation method that is specifically designed for optimizing \ac{PL}-ranking models w.r.t.\ ranking metrics.
While both the placement policy gradient estimator (Eq.~\ref{eq:placementapproximation}) and PL-Rank-1 (Eq.~\ref{eq:plrank1}) estimate the same gradient, their formulas look radically different.
A big advantage of PL-Rank-1 is that it can be computed with a time-complexity of
$
\mathcal{O}( N \cdot K \cdot D)
$.
Our experimental results indicate that while both estimators have comparable sample-efficiency, PL-Rank-1 requires considerably less time to compute than using a machine learning framework to automatically compute the placement policy gradient.

\subsection{Improving Sample-Efficiency} %

In Eq.~\ref{eq:plrank1fullderiv} we see that an item receives a positive weight from the expected following reward.
Therefore, even when an item has a low probability of being placed it can compensate with a high relevance ($\rho_d$) to get a positive weight.
However, when an estimate of the gradient is based on a low number of samples ($N$), they may not include a ranking where such an item is placed at all and thus these items will nevertheless receive a negative weight in the estimate.
We propose one last estimator to mitigate this potential issue.

First, we can rewrite the expected reward following placement so that the reward obtained from $d$ and that from items placed afterwards are separated:
\begin{equation}
\begin{split}
\mathbb{E}_{y}
\hspace{-0.26cm}&\hspace{0.26cm}
\mleft[
\sum_{k=\text{rank}(d, y)}^K
 \theta_k
\rho_{y_k}
\mright]
\\&=
\mathbb{E}_{y} \mleft[
\mleft(\sum_{k=\text{rank}(d, y) + 1}^K
 \theta_k
\rho_{y_k}
\mright)
+
 \theta_{\text{rank}(d, y)}
\rho_d
\mright]
\\&=
\mathbb{E}_{y} \mleft[
\mleft(\sum_{k=\text{rank}(d, y) + 1}^K
 \theta_k
\rho_{y_k}
\mright)
+
\sum_{k=1}^K
\pi(d \mid y_{1:k-1})
 \theta_k
\rho_d
\mright]
\\&=
\mathbb{E}_{y} \mleft[
\mleft(\sum_{k=\text{rank}(d, y) + 1}^K
 \theta_k
\rho_{y_k}
\mright)
+
\sum_{k=1}^{\text{rank}(d, y)}
\pi(d \mid y_{1:k-1})
 \theta_k
\rho_d
\mright],
\end{split}
\end{equation}
where again we make use of the fact that: $k > \text{rank}(d, y) \rightarrow \pi(d \mid y_{1:k-1}) = 0$.
Combining this result with Eq.~\ref{eq:plrank1fullderiv} we get:
\begin{equation}
\begin{split}
\frac{\delta}{\delta m} 
\mathcal{R}(q)
&=
\sum_{d \in \mathcal{D}} \mleft[\frac{\delta}{\delta m} m(d) \mright]
\mathbb{E}_{y} \Bigg[
\overbrace{
\mleft(\sum_{k=\text{rank}(d, y)+1}^K
 \theta_k
\rho_{y_k}
\mright)
}^\text{future reward after placement}
\\
&\quad\,\,\,\,
+
\underbrace{
\sum_{k=1}^{\text{rank}(d, y)}
 \pi(d \mid y_{1:k-1}) 
\mleft( \theta_k
\rho_d - \sum_{x=k}^K \theta_x
\rho_{y_x}
\mright)
}_\text{expected direct reward minus the risk of placement}
\Bigg].
\end{split}
\label{eq:plrank2fullderiv}
\end{equation}
We see that the gradient w.r.t.\  an item's log score $m(d)$ is weighted by the reward after placement (not including the reward from $d$) plus the expected direct reward (the reward from $d$) minus the expected risk imposed by $d$ before its placement.
From Eq.~\ref{eq:plrank2fullderiv} we can derive the following novel estimator:
\begin{equation}
\begin{split}
\frac{\delta}{\delta m} 
\mathcal{R}(q)
&\approx
\frac{1}{N}
\sum_{d \in \mathcal{D}} \mleft[\frac{\delta}{\delta m} m(d) \mright]
\sum_{i=1}^N
\mleft(\sum_{k=\text{rank}(d, y^{(i)})+1}^K
 \theta_k
\rho_{y_k^{(i)}}
\mright)
\\
&\hspace{0.8cm}
+
\sum_{k=1}^{\text{rank}(d, y^{(i)})}
 \pi(d \mid y_{1:k-1}^{(i)}) 
\mleft( \theta_k
\rho_d - \sum_{x=k}^K \theta_x
\rho_{y_x^{(i)}}
\mright)
.
\end{split}
\label{eq:plrank2}
\end{equation}
We will call this estimator: \emph{PL-Rank-2}.
Unlike PL-Rank-1 (Eq.~\ref{eq:plrank1}), PL-Rank-2 can provide a positive weight to items that were not in the top-K of any of the $N$ sampled rankings.
While this is expected to increase the sample-efficiency, it does not come at the cost of computational complexity as both PL-Rank-1 and PL-Rank-2 have a complexity of
$
\mathcal{O}( N \cdot K \cdot D)
$.

\subsection{The PL-Rank Algorithm}

Finally, we will show how PL-Rank-2 can be implemented efficiently.
Our goal is to compute a $\lambda_d$ weight per item $d$ so that the gradient is estimated by:
\begin{equation}
\frac{\delta}{\delta m} 
\mathcal{R}(q)
\approx
\frac{1}{N}
\sum_{d \in \mathcal{D}} \lambda_d \mleft[\frac{\delta}{\delta m} m(d) \mright],
\end{equation}
where following Eq.~\ref{eq:plrank2} these weights are:
\begin{equation}
\begin{split}
\lambda_d = \frac{1}{N} \sum_{i=1}^N &
\Bigg(
\mleft(\sum_{k=\text{rank}(d, y)+1}^K
 \theta_k
\rho_{y_k}
\mright)
\\ 
&\,\, +
\sum_{k=1}^{\text{rank}(d, y)}
 \pi(d \mid y_{1:k-1}) 
\mleft( \theta_k
\rho_d - \sum_{x=k}^K \theta_x
\rho_{y_x}
\mright)
\Bigg).
\end{split}
\end{equation}
Algorithm~\ref{alg:plrank2} displays the PL-Rank-2 algorithm in pseudo-code.
As input it requires the item collection $\mathcal{D}$, the relevances $\rho$, the (pre-computed) log scores per item $m(d)$ according to the current model $m$, the metric weights per rank $\theta$, and finally, the number of rankings to sample $N$ (Line~\ref{alg:line:input}).
First, $N$ rankings are sampled using Gumbel sampling (Line~\ref{alg:line:gumbell}) and zero weights are initialized for every $\lambda$ (Line~\ref{alg:line:gumbell}).
Subsequently, the initial denominator for the \ac{PL} model is computed and stored (Line~\ref{alg:line:initdenom}).
Then the algorithm starts iterating over each of the sampled rankings, where first, the rewards following each rank $k$ are precomputed (Line~\ref{alg:line:cumreward}).
Second, it loops over every rank $k$ where it adds the following reward to the $\lambda_d$ of $d$ at rank $k$ in the sampled ranking (Line~\ref{alg:line:futurereward}), thus computing the first part of Eq.~\ref{eq:plrank2}.
For every item, the placement probability $\pi(d \mid y_k^{(i)})$ is computed (Line~\ref{alg:line:placeprob}) and multiplied by the difference between the item's direct reward and the following reward (Line~\ref{alg:line:2ndpart}), this is added to $\lambda_d$ to compute the second part of Eq.~\ref{eq:plrank2}.
Finally, the denominator is updated to account for the item placed at rank $k$ (Line~\ref{alg:line:denomupdate}).

Algorithm~\ref{alg:plrank2} reveals that PL-Rank-2 can be computed in
$
\mathcal{O}( N \cdot K \cdot D)
$, we note that with small alterations to Line~\ref{alg:line:futurereward}~and~\ref{alg:line:2ndpart} PL-Rank-1 can be computed with this algorithm as well.

\begin{algorithm}[t]
\caption{PL-Rank-2 Gradient Estimation} 
\label{alg:plrank2}
\begin{algorithmic}[1]
\STATE \textbf{Input}: items: $\mathcal{D}$; Relevances: $\rho$; Metric weights: $\theta$;
\\\phantom{ \textbf{Input}}
Scores: $m$; Number of samples: $N$. \label{alg:line:input}
\STATE $\{y^{(1)}, y^{(2)},\dots,y^{(N)}\} \leftarrow \text{Gumbel\_Sample}(N, m)$ \label{alg:line:gumbell}
\STATE $\lambda \leftarrow \mathbf{0}$ \hfill \textit{\small // initialize zero weight per item}  \label{alg:line:initlambda}
\STATE $M \leftarrow \sum_{d\in\mathcal{D}} \exp(m(d))$ \hfill \textit{\small // initialize PL denominator}  \label{alg:line:initdenom}
\FOR{$i \in [1,2,\ldots,N]$}
    \STATE $M' \leftarrow M$ \hfill \textit{\small // copy initial PL denominator} 
    \STATE $\omega_{K} \leftarrow \theta_K\rho_{y_K^{(i)}}$  \hfill \textit{\small // reward for last rank}
    \FOR{$k \in [K-1, K-2, \ldots,1]$}
    \STATE $\omega_k \leftarrow \omega_{k+1} + \theta_k\rho_{y_k^{(i)}}$  \hfill \textit{\small // pre-compute reward following rank $k$} \label{alg:line:cumreward}
    \ENDFOR
    \FOR{$k \in [1, 2,\ldots, K]$}
    \STATE $\lambda_{y_k^{(i)}} \leftarrow \lambda_{y_k^{(i)}} + \omega_{k+1}$ \hfill \textit{\small // add future reward $k$} \label{alg:line:futurereward}
        \FOR{$d \in \mathcal{D}$}
            \IF{$d \not\in y_{1:k-1}^{(i)}$}
            \STATE $P \leftarrow  (\exp(m(d))/M')$ \hfill \textit{\small // placement probability} \label{alg:line:placeprob}
            \STATE $\lambda_d \leftarrow \lambda_d + P \cdot (\theta_k\rho_d - \omega_{k})$ \hfill \textit{\small // 2nd part of Eq.~\ref{eq:plrank2}} \label{alg:line:2ndpart}
            \ENDIF
        \ENDFOR
    \STATE $M' \leftarrow M' - \exp(m(y_k^{(i)}))$  \hfill \textit{\small // renormalize denominator} \label{alg:line:denomupdate}
    \ENDFOR
\ENDFOR
\RETURN $\frac{1}{N}\lambda$ \label{alg:line:return}
\end{algorithmic}
\end{algorithm}

\section{Method: PL-Rank for Fairness}
\label{sec:fairness}

So far we have introduced the PL-Rank algorithms for estimating the gradient of a \ac{PL} ranking model w.r.t.\ a relevance metric.
However, the applicability of these algorithms are much wider than just relevance metrics, in particular, they can be applied to any \emph{exposure-based} metrics~\citep{singh2019policy, diazevaluatingstochastic, morik2020, biega2018equity}.
Exposure represents the expected number of people that will examine an item.
In general, user behavior has \emph{position-bias} which means that they are less likely to examine an item if it displayed at a lower rank~\citep{wang2018position, craswell2008experimental}.
Let the rank weight $\theta_k$ indicate the probability that a user examines an item at rank $k$, then the exposure an item $d$ receives under $\pi$ is:
\begin{equation}
 \mathcal{E}(q, d)
 = \mathbb{E}_y\mleft[\sum_{k=1}^K \theta_k \mathds{1}[y_k = d]\mright]
 = \sum_{y \in \pi} \pi(y) \sum_{k=1}^K \theta_k \mathds{1}[y_k = d],
 \label{eq:exposure}
\end{equation}
where again for brevity we denote $\mathcal{E}_d = \mathcal{E}(q, d)$.
Thus $\mathcal{E}_d$ could be interpreted as the probability that a user examines $d$ when $\pi$ is deployed.
Most fairness metrics for rankings consider how exposure is distributed over items and  specifically how \emph{fair} this distribution is.
Regardless of the exact metric, PL-Rank can be applied to a fairness metric $\mathcal{F}$ if the chain-rule can be applied as follows:
\begin{equation}
\frac{\delta}{\delta m}
\mathcal{F}(q) =
\sum_{d \in \mathcal{D}}
\frac{\delta \mathcal{F}(q)}{\delta \mathcal{E}_d}
\frac{\delta \mathcal{E}_d}{\delta m}.
\label{eq:fairgrad1}
\end{equation}
To derive this gradient, we first note that the $\mathcal{E}_d$ (Eq.~\ref{eq:exposure}) and $\mathcal{R}(q)$ (Eq.~\ref{eq:queryreward}) are equivalent if $\forall d'\, \rho_{d'} = \mathds{1}[d' = d]$, therefore if we replace $\rho$ in PL-Rank-2 (Eq.~\ref{eq:plrank2fullderiv}) accordingly, it will provide us the gradient $\frac{\delta \mathcal{E}_d}{\delta m}$.
If we combine this fact with Eq.~\ref{eq:fairgrad1} we obtain the following PL-Rank-2 based gradient:
\begin{align}
\frac{\delta}{\delta m}
\mathcal{F}&(q)
=
\sum_{d \in \mathcal{D}} \mleft[\frac{\delta}{\delta m} m(d) \mright]
\mathbb{E}_{y} \Bigg[
\mleft(\sum_{k=\text{rank}(d, y)+1}^K
 \theta_k
\mleft[ \frac{\delta \mathcal{F}(q)}{\delta \mathcal{E}_{y_k}} \mright]
\mright)
\\
&
+
\sum_{k=1}^{\text{rank}(d, y)}
 \pi(d \,|\, y_{1:k-1}) 
\mleft( \theta_k
\mleft[ \frac{\delta \mathcal{F}(q)}{\delta \mathcal{E}_d} \mright]
 - \sum_{x=k}^K \theta_x
\mleft[ \frac{\delta \mathcal{F}(q)}{\delta \mathcal{E}_{y_x}} \mright]
\mright)
\Bigg].
\nonumber
\end{align}
In other words, we can apply PL-Rank by simply replacing the item relevances with the gradients: $\frac{\delta \mathcal{F}(q)}{\delta \mathcal{E}_d}$ before computation.
Similarly, any linear combination of $\mathcal{R}$ and $\mathcal{F}$ can be optimized by replacing the relevances with the corresponding linear combination between $\rho_d$ and $\frac{\delta \mathcal{F}(q)}{\delta \mathcal{E}_d}$.

For instance, we can follow \citet{singh2019policy} and choose a disparity-based metric.
This metric measures the disparity between two items via a function $D(d, d')$ and takes the average disparity over all item pairs:
\begin{equation}
\mathcal{F}(q) = 
\frac{1}{|\mathcal{D}|(|\mathcal{D}| - 1)}
\sum_{d \in \mathcal{D}} \sum_{d' \in \mathcal{D}}  D(d, d').
\end{equation}
\citet{singh2019policy} divide the exposure of an item $d$ by its relevance: $\frac{\mathcal{E}_d}{\rho_d}$ to model the proportion between the exposure and the merit of an item.
However, in our experimental datasets many items have zero relevances, thus making such a division impossible.
Instead, we introduce a novel alternative disparity measure:
\begin{equation}
D(d, d') = \mleft(
\mathcal{E}_{d'} \rho_{d} -  \mathcal{E}_d \rho_{d'}
\mright)^2.
\end{equation}
This measure looks at the reward item $d$ would receive if it had the exposure of $d'$: $\mathcal{E}_{d'} \rho_{d}$, in other words, the reward $d$ would receive if it was treated as $d'$ is.
This measure can handle items without merit and has the gradient:
\begin{equation}
\frac{\delta \mathcal{F}(q)}{\delta \mathcal{E}_d}
= \frac{4}{|\mathcal{D}|(|\mathcal{D}| - 1)}
\sum_{d' \in \mathcal{D}} 
\mleft(
\mathcal{E}_{d'} \rho_{d} -  \mathcal{E}_d \rho_{d'}
\mright) \rho_{d'}.
\label{eq:fairnessgradexposure}
\end{equation}
Thus in order to apply PL-Rank-2 to this fairness metric, one only needs to compute (or estimate) Eq.~\ref{eq:fairnessgradexposure} and then run Algorithm~\ref{alg:plrank2} where the relevances are replaced with the gradients: $\frac{\delta \mathcal{F}(q)}{\delta \mathcal{E}_d}$.

To conclude, we have shown that PL-Rank-2 can efficiently estimate the gradient of exposure-based fairness metrics, in addition to relevance metrics and any linear combination of any set of these metrics.

\section{Experimental Setup}
\label{sec:experimentalsetup}

The experiments performed for this paper aim to answer three research questions:
\begin{enumerate}[align=left, label={\bf RQ\arabic*},leftmargin=*]
\item Does PL-Rank require fewer sampled rankings for optimal convergence than policy gradients or LambdaLoss?
\label{rq:sample}
\item Is less computational time needed to reach high performance with PL-Rank than with policy gradients or LambdaLoss?\label{rq:time}
\item Is PL-Rank also effective at optimizing an exposure-based fairness metric?
\label{rq:fair}
\end{enumerate}
In other words, we address the sample-efficiency and the computational costs of PL-Rank, in addition to its applicability to ranking-fairness metrics.

To evaluate these aspects we compare with three baselines:
\begin{enumerate*}[label=(\roman*)]
\item the policy gradient as described in Section~\ref{sec:basicpolicygradient},
this is the most basic form of gradient estimation~\citep{singh2019policy, williams1992simple, bruch2020stochastic};
\item the placement policy gradient as introduced in Section~\ref{sec:placementpolicygradient}, this gradient estimation considers individual item placements;
and \item LambdaLoss~\citep{wang2018lambdaloss}, a state-of-the-art heuristic for optimizing deterministic ranking models.
Following \citet{bruch2020stochastic} we apply an average of the gradients over $N$ sampled rankings.
One can easily extend the existing proof that LambdaLoss optimizes a lower bound on the performance of a deterministic model~\citep{wang2018lambdaloss} to prove our approach also optimizes a lower bound on the expected performance of a stochastic \ac{PL} ranking model.
\end{enumerate*}
We note that \citet{bruch2020stochastic} introduced additional heuristic methods for PL-Ranking model optimization, due to their high similarity with LambdaLoss we omitted these methods from our baselines.
To the best of our knowledge, our choice of baselines cover every category of existing methods for the metric-based optimization of PL-Ranking models.

We base our experiments on the three largest publicly-available \ac{LTR} industry datasets: \emph{Yahoo!\ Webscope}~\citep{Chapelle2011}, \emph{MSLR-WEB30k}~\citep{qin2013introducing}, and \emph{Istella}~\citep{dato2016fast}.
Each dataset contains queries, preselected documents per query, and relevance labels indicating the expert-judged relevance of a preselected document w.r.t.\ a query.
Query-document combinations are represented by feature vectors, each dataset varies in the number of features, queries and average number of preselected documents:
Yahoo\ contains 29,921 queries and on average 24 preselected documents per query encoded in 700 features; MSLR has 30,000 queries, on average 125 documents per query and 136 features; and lastly, Istella has 33,118 queries, on average 315 documents per query and 220 features.

For our relevance experiments, we optimize top-5 \emph{Discounted Cumulative Gain} (DCG@5) and  choose $\theta$ accordingly (Eq.~\ref{eq:dcg@5}).\footnote{We chose DCG instead of normalized DCG based on the advice of \citet{ferrante2021towards}.}
The relevance of a document is set to a transformation of its label: $\rho_d = 2^{\text{relevance\_label}(d)} - 1$.
For the fairness experiments, we optimize the disparity metric introduced in Section~\ref{sec:fairness}, exposure values $\mathcal{E}_d$ are estimated using $1000$ sampled rankings.
To compare the computational costs of each method, we ran repeated experiments under identical circumstances on a single \emph{Intel Xeon Silver 4214} CPU and measured the time taken to complete each epoch.
All our reported results are averaged over 20 independent runs.

Based on preliminary parameter tuning, we chose to optimize neural networks with two hidden layers of 32 sigmoid activated nodes, we used standard stochastic gradient descent with a $0.01$ learning rate for all methods.
For calculating gradients we utilize \emph{Tensorflow}~\citep{abadi2016tensorflow} with two exceptions: the sampling of rankings and $\frac{\delta \mathcal{R}(q)}{\delta m}$ with the PL-Rank algorithm (Algorithm~\ref{alg:plrank2}) are computed using \emph{Numpy}~\citep{harris2020array}.

\setlength{\tabcolsep}{2pt}
\begin{table}[tb]
\caption{
Average time in minutes taken to perform one training epoch for different numbers of sampled rankings $N$, the standard deviation is displayed in brackets.
}
\begin{tabular}{c l c c c c}
&
& $N=1$
& $N=10$
& $N=100$
& $N=1000$
\\
\hline
\multirow{5}{*}{\rotatebox[origin=lt]{90}{\small Yahoo}}
& \small LambdaLoss
& 2.48 {\tiny ( 0.05)}
& 2.53 {\tiny ( 0.04)}
& 3.06 {\tiny ( 0.08)}
& 10.25 {\tiny ( 0.53)}
\\
& \small Policy Gradient
& 3.79 {\tiny ( 0.09)}
& 3.80 {\tiny ( 0.06)}
& 4.28 {\tiny ( 0.15)}
& 8.27 {\tiny ( 0.50)}
\\
& \small Placement P.G.
& 3.83 {\tiny ( 0.08)}
& 3.86 {\tiny ( 0.05)}
& 4.42 {\tiny ( 0.10)}
& 8.26 {\tiny ( 0.44)}
\\
& \small PL-Rank-1
& \bf 2.45 {\tiny ( 0.06)}
& \bf 2.49 {\tiny ( 0.06)}
& \bf 2.82 {\tiny ( 0.09)}
& \bf 5.70 {\tiny ( 0.14)}
\\
& \small PL-Rank-2
& 2.49 {\tiny ( 0.06)}
& 2.52 {\tiny ( 0.06)}
& 2.87 {\tiny ( 0.08)}
& 6.22 {\tiny ( 0.15)}
\\
\hline
\multirow{5}{*}{\rotatebox[origin=lt]{90}{\small MSLR}}
& \small LambdaLoss
& 2.73 {\tiny ( 0.11)}
& 3.96 {\tiny ( 0.59)}
& 36.36 {\tiny ( 31.46)}
& 1669.59 {\tiny ( 450.69)}
\\
& \small Policy Gradient
& 3.30 {\tiny ( 0.10)}
& 3.45 {\tiny ( 0.10)}
& 5.25 {\tiny ( 0.35)}
& 24.20 {\tiny ( 2.77)}
\\
& \small Placement P.G.
& 3.32 {\tiny ( 0.17)}
& 3.42 {\tiny ( 0.13)}
& 5.27 {\tiny ( 0.41)}
& 23.97 {\tiny ( 2.68)}
\\
& \small PL-Rank-1
& \bf 2.16 {\tiny ( 0.14)}
& \bf 2.28 {\tiny ( 0.13)}
& \bf 3.34 {\tiny ( 0.13)}
& \bf 18.48 {\tiny ( 2.10)}
\\
& \small PL-Rank-2
& 2.19 {\tiny ( 0.15)}
& 2.35 {\tiny ( 0.17)}
& 3.45 {\tiny ( 0.06)}
& 21.10 {\tiny ( 2.78)}
\\
\hline
\multirow{5}{*}{\rotatebox[origin=lt]{90}{\small Istella}}
& \small LambdaLoss
& 3.53 {\tiny ( 0.12)}
& 4.50 {\tiny ( 0.10)}
& 27.81 {\tiny ( 19.20)}
& 142.74 {\tiny ( 16.25)}
\\
& \small Policy Gradient
& 4.10 {\tiny ( 0.17)}
& 4.51 {\tiny ( 0.16)}
& 8.74 {\tiny ( 0.26)}
& 44.29 {\tiny ( 2.17)}
\\
& \small Placement P.G.
& 4.08 {\tiny ( 0.17)}
& 4.51 {\tiny ( 0.18)}
& 8.72 {\tiny ( 0.23)}
& 44.74 {\tiny ( 2.55)}
\\
& \small PL-Rank-1
& \bf 3.01 {\tiny ( 0.12)}
& \bf 3.27 {\tiny ( 0.09)}
& \bf 6.90 {\tiny ( 0.19)}
& \bf 39.80 {\tiny ( 2.65)}
\\
& \small PL-Rank-2
& 3.04 {\tiny ( 0.13)}
& 3.31 {\tiny ( 0.10)}
& 7.02 {\tiny ( 0.14)}
& 40.64 {\tiny ( 3.96)}
\\
\hline
\end{tabular}
\label{tab:time}
\end{table}

{\renewcommand{\arraystretch}{0.03}
\begin{figure*}[tb]
\centering
\begin{tabular}{c r r r}
&
 \multicolumn{1}{c}{ \small  $N = 10$}
&
 \multicolumn{1}{c}{ \small $N = 100$}
&
 \multicolumn{1}{c}{ \small $N = 1000$}
\\
\rotatebox[origin=lt]{90}{\hspace{0.35em} \small Yahoo! Webscope} &
\includegraphics[scale=0.36]{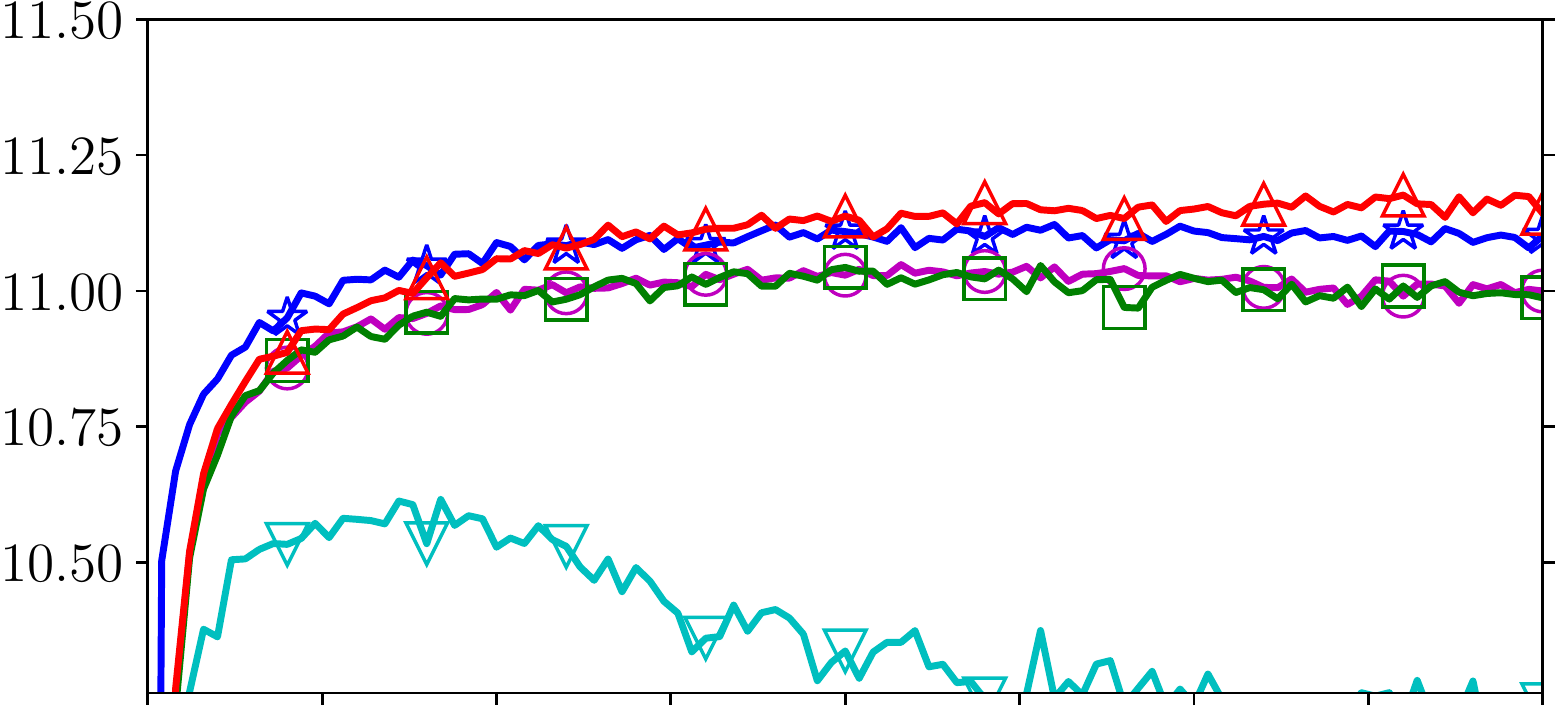}\hspace{0.6mm} &
\includegraphics[scale=0.36]{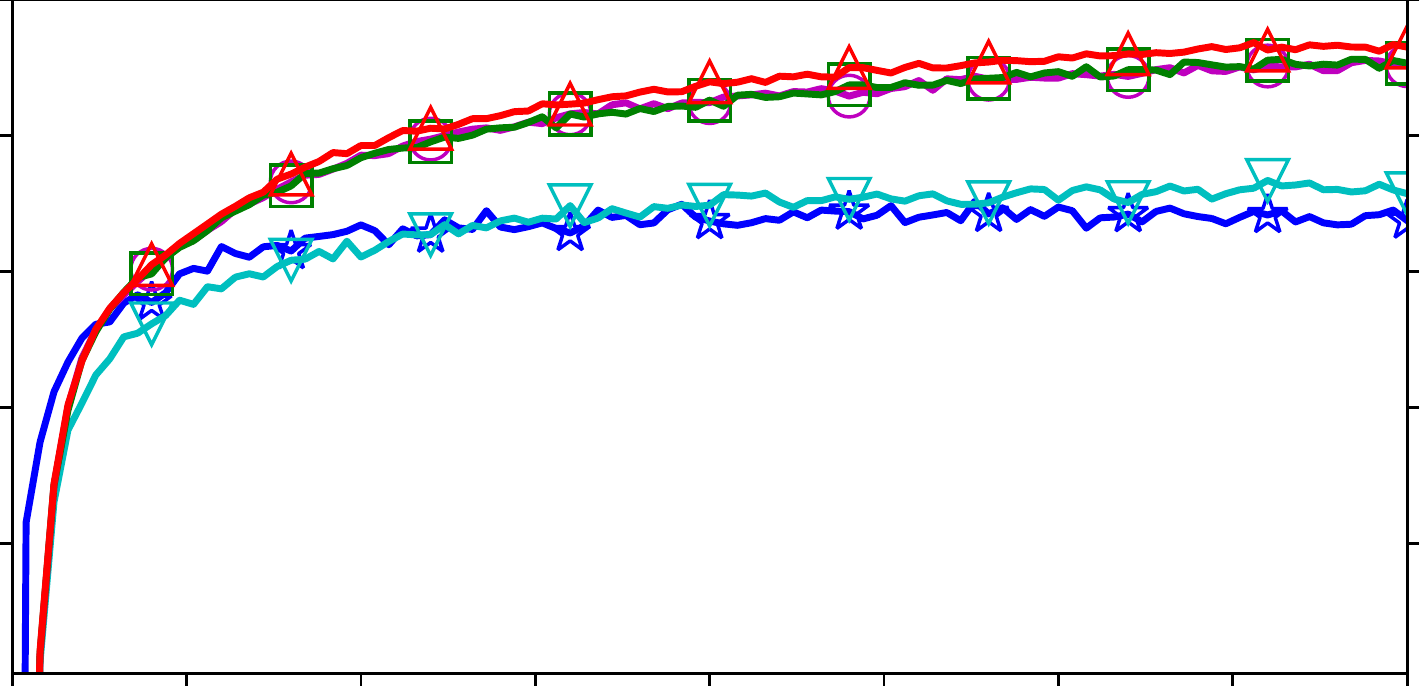}\hspace{0.55mm} &
\includegraphics[scale=0.36]{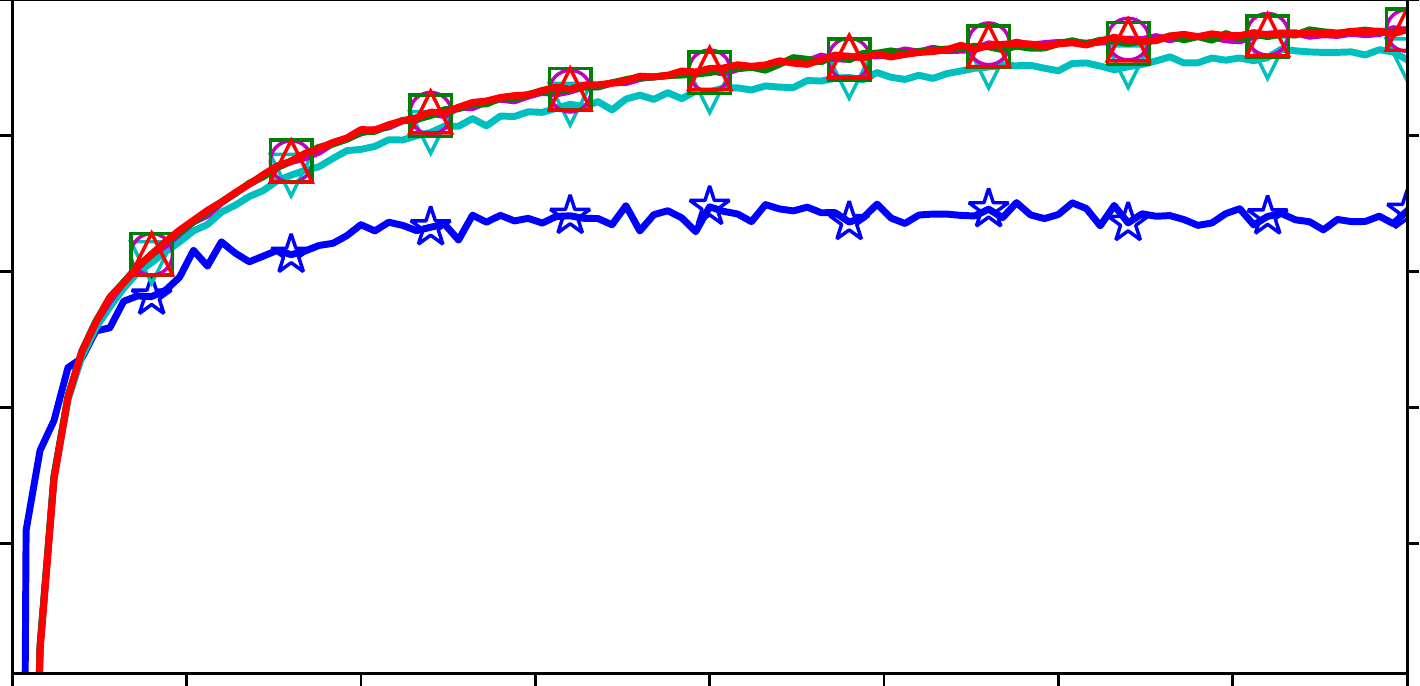}\hspace{0.6mm} 
\\
\rotatebox[origin=lt]{90}{\hspace{0.8em} \small MSLR-WEB30k} &
\includegraphics[scale=0.36]{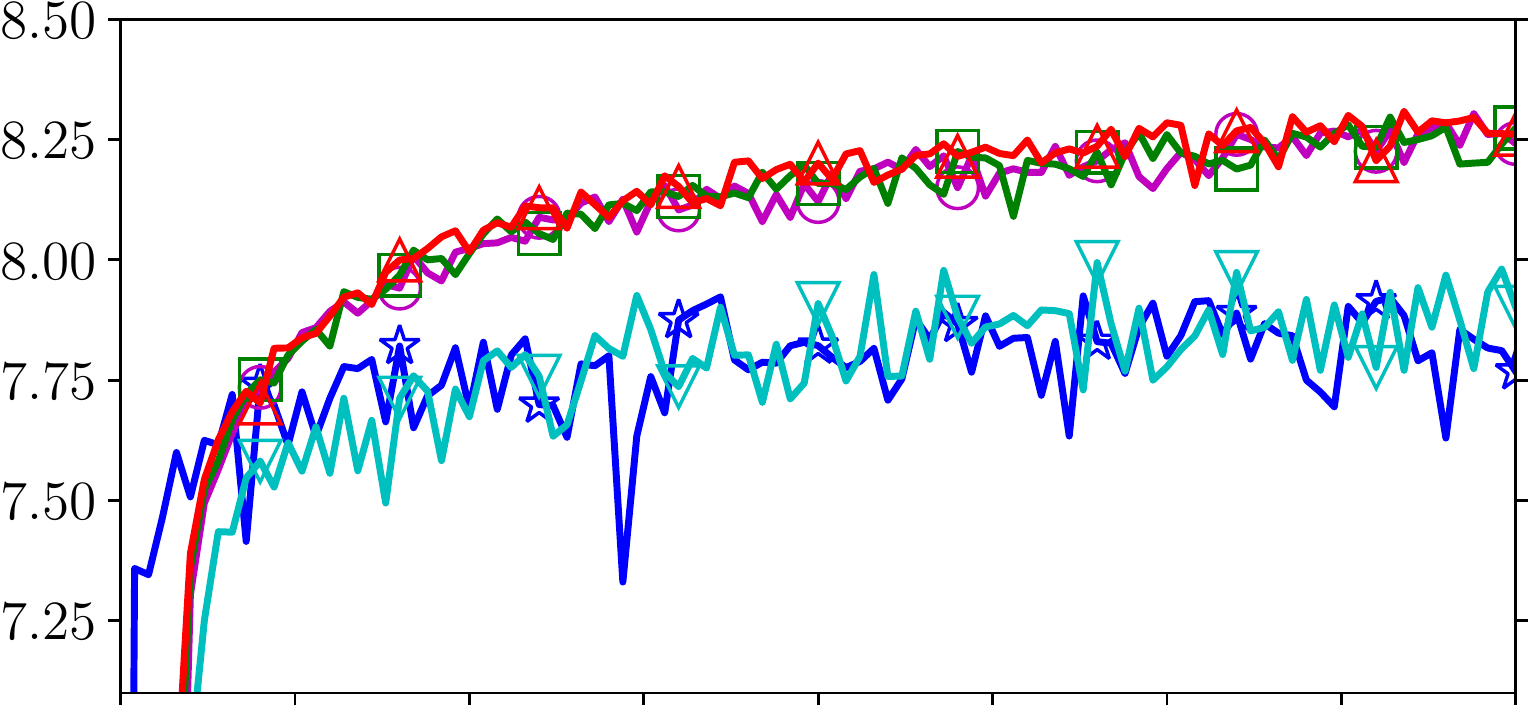}\hspace{0.6mm} &
\includegraphics[scale=0.36]{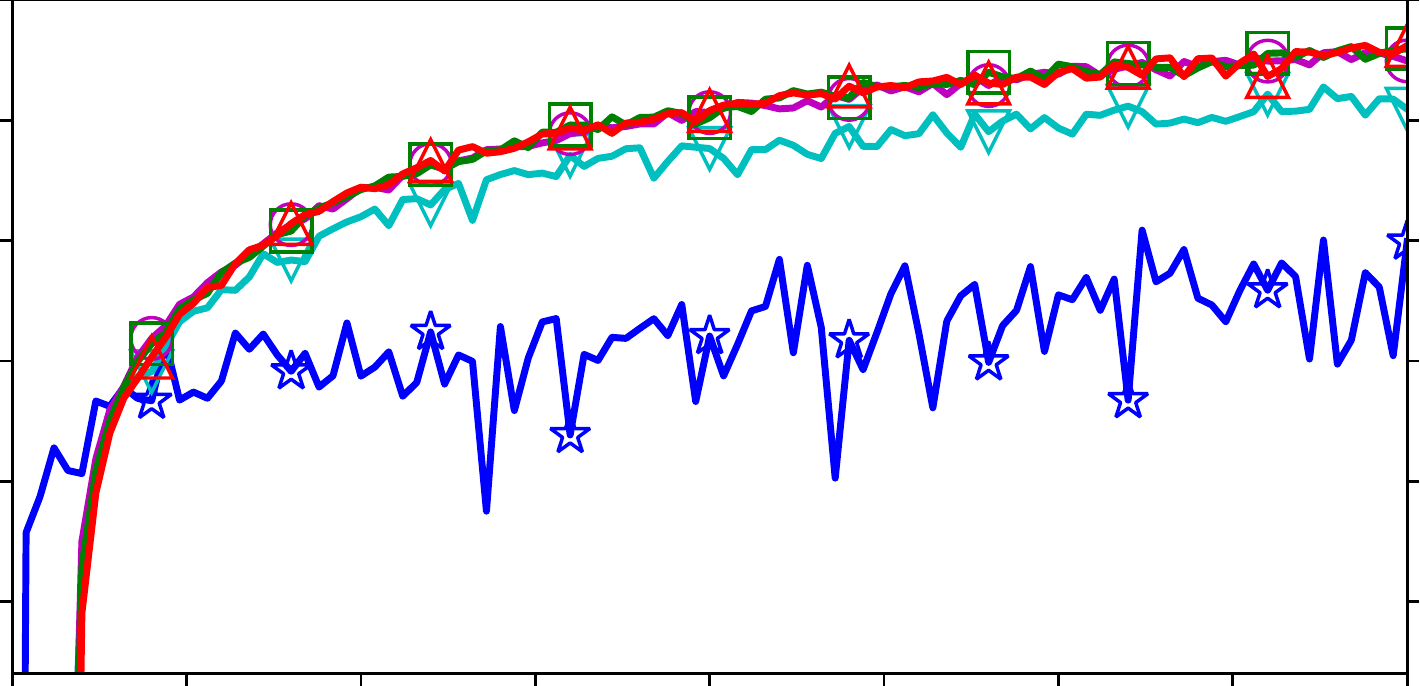}\hspace{0.55mm} &
\includegraphics[scale=0.36]{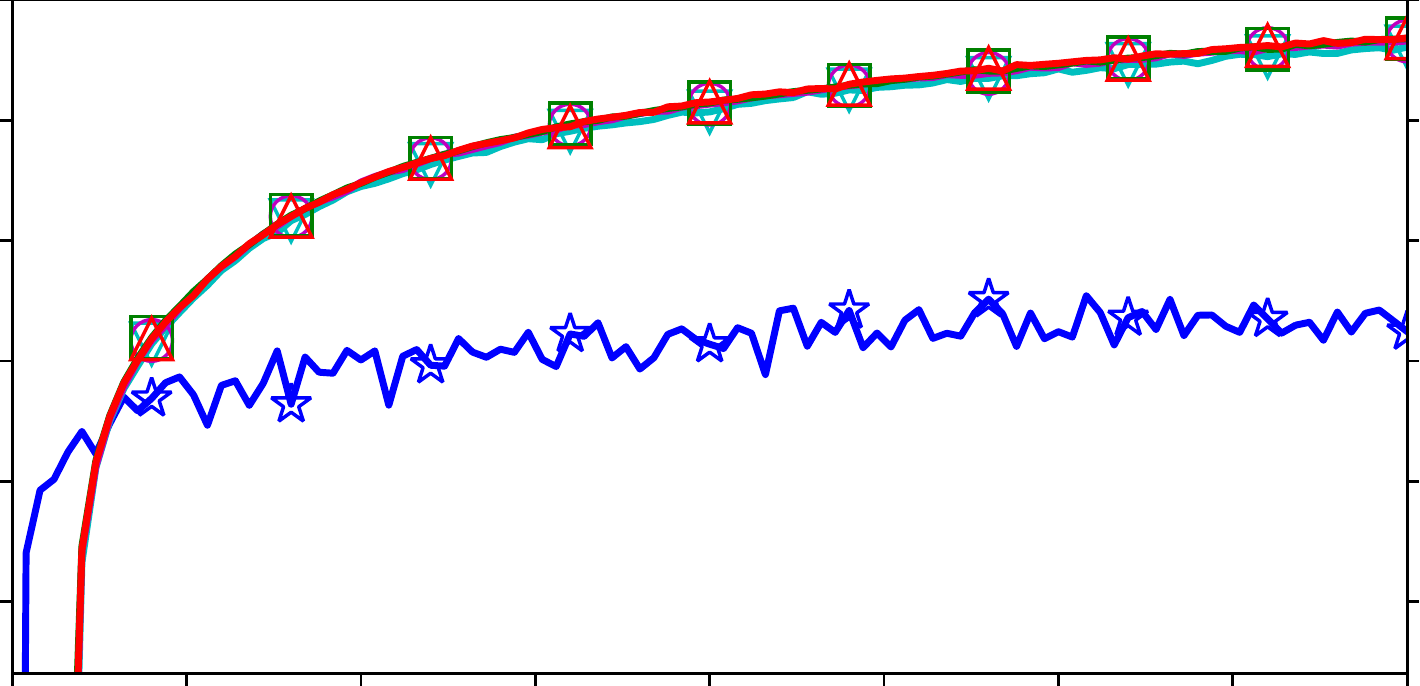}\hspace{0.6mm} 
\\
\rotatebox[origin=lt]{90}{\hspace{2.8em} \small Istella} &
\includegraphics[scale=0.36]{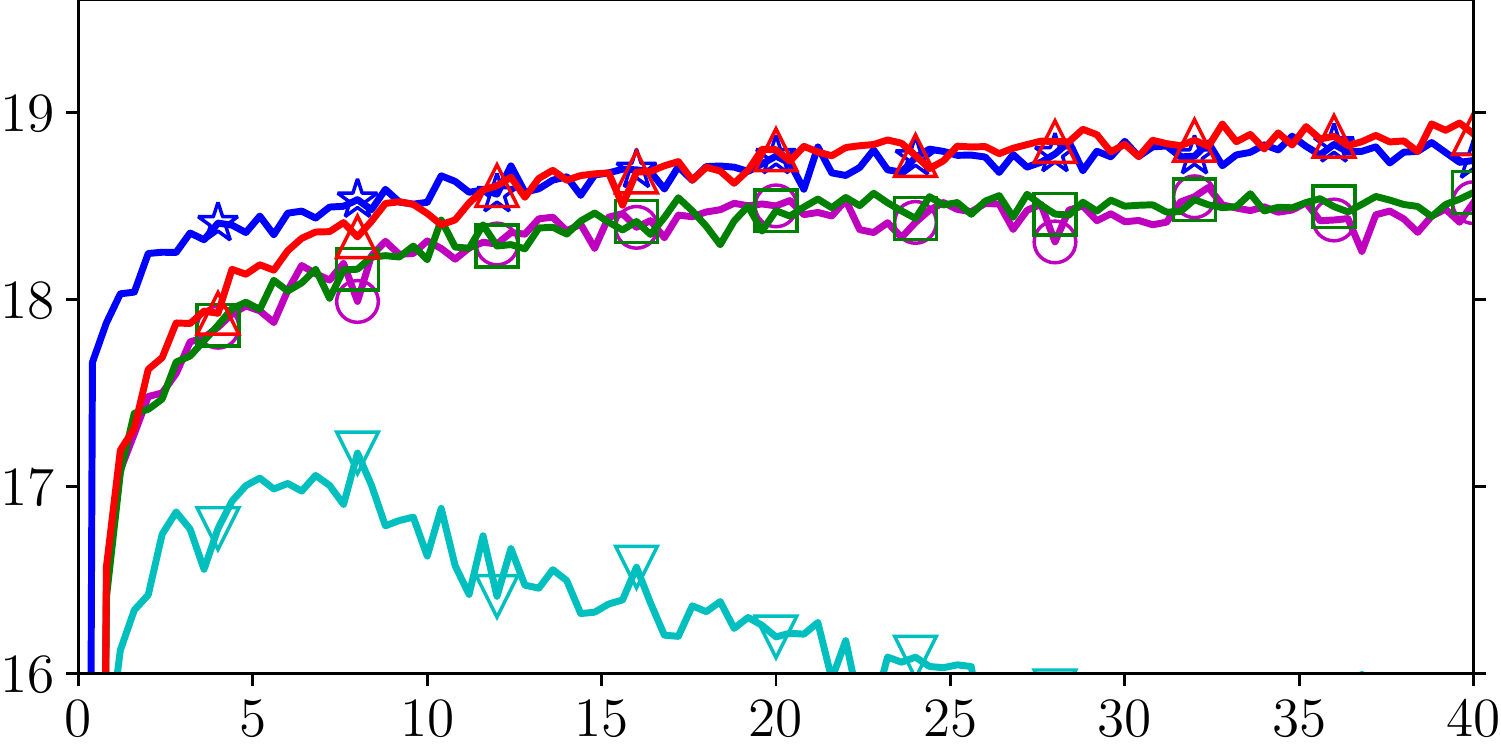} &
\includegraphics[scale=0.36]{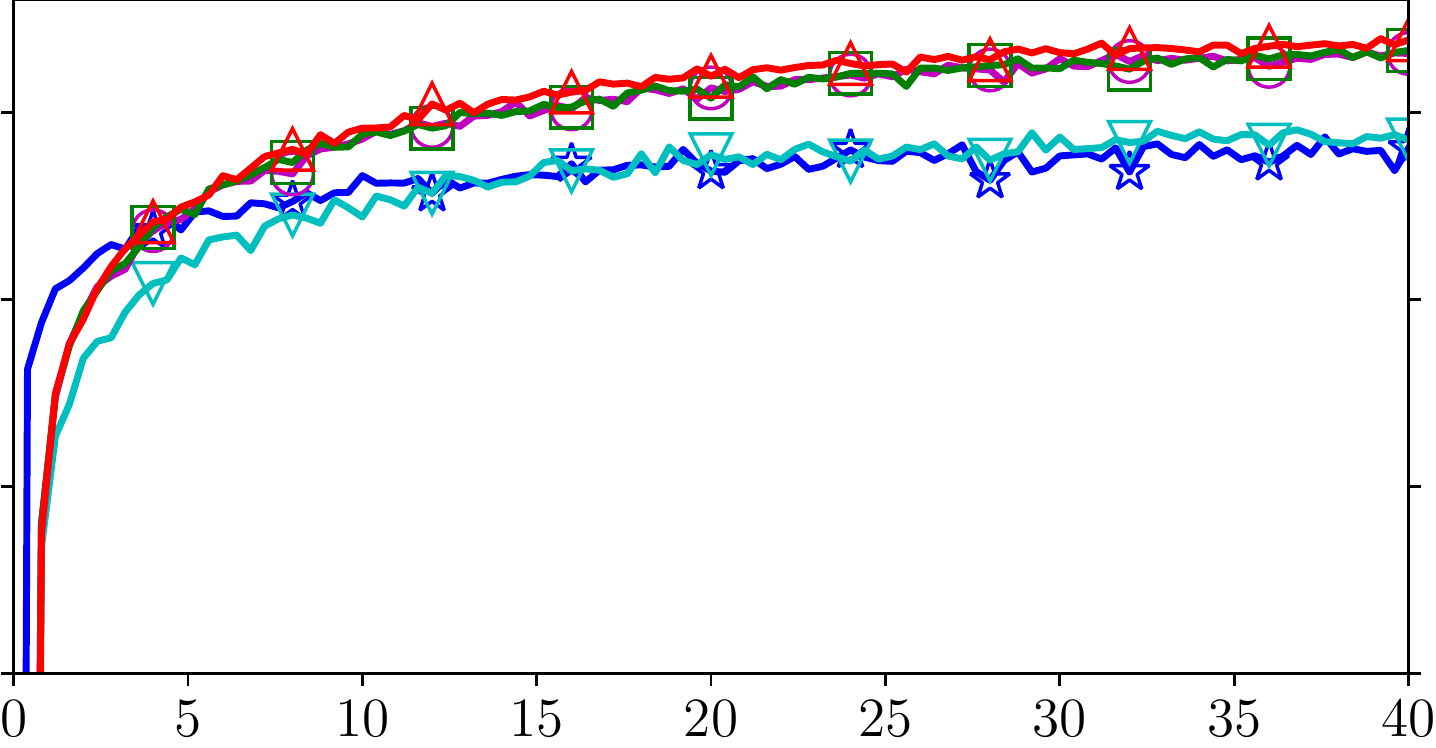} &
\includegraphics[scale=0.36]{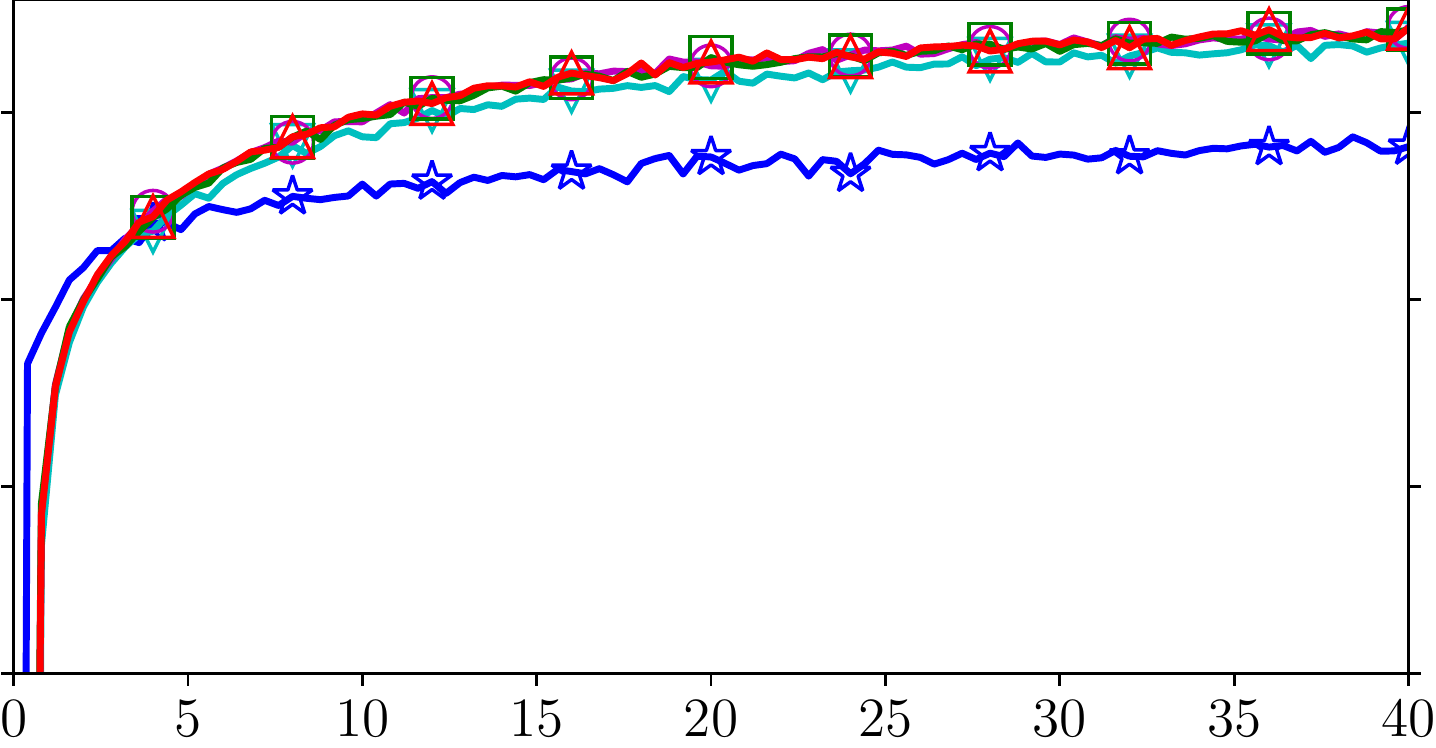} 
\\
& \multicolumn{1}{c}{\small \hspace{0.5em} Training Epoch}
& \multicolumn{1}{c}{\small \hspace{0.5em} Training Epoch}
& \multicolumn{1}{c}{\small \hspace{0.5em} Training Epoch}
\\
\multicolumn{4}{c}{
\includegraphics[scale=.45]{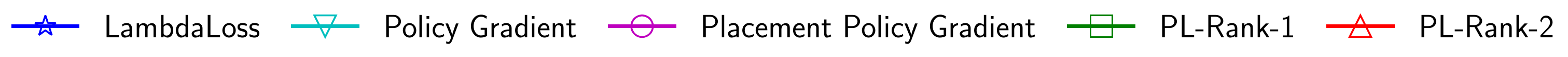}
} 
\end{tabular}
\caption{
Performance in DCG@5 of PL ranking models trained using different gradient estimation methods with varying number of sampled rankings $N$ per estimation.
Results are the mean of 20 independent runs.
}
\vspace{-0.3\baselineskip}
\label{fig:sample}
\end{figure*}
}

\section{Results}
\label{sec:results}
Our discussion of the results is divided per research question, our results are displayed in Figure~\ref{fig:sample},~\ref{fig:dynamic} and~\ref{fig:fairness} and Table~\ref{tab:time} and~\ref{tab:dcg}.

\subsection{Sample-Efficiency}

We will first consider \ref{rq:sample}: \emph{whether PL-Rank needs fewer sampled rankings for optimal convergence}.
Figure~\ref{fig:sample} shows the performance of PL ranking models trained using different gradient estimation methods with varying numbers of sampled rankings used for estimation.
We see that increasing the number of samples beyond $N = 10$ does not have any noticeable effect on the performance of LambdaLoss.
In all cases, LambdaLoss converges at suboptimal performance after only a few epochs.
In contrast, the basic policy gradient is very affected by $N$ and on all three datasets it requires $N=1000$ to get close to optimal performance, it has extreme convergence issues when $N=10$.
However, in all cases the placement policy gradient outperforms the basic policy gradient and can converge near optimal performance with $N=100$.
The performance of PL-Rank-1 and the placement policy gradient are indistinguishable. %
This strongly suggests that PL-Rank-1 and the placement policy gradient perform the same estimation, although Section~\ref{sec:resultstime} will reveal that PL-Rank-1 does so in a more computationally efficient way.
Lastly, we see that PL-Rank-2 outperforms PL-Rank-1 and the policy gradient methods when $N=10$ on the Yahoo and Istella datasets.
Noticeable but limited improvements are also present with $N=100$ on these datasets.
Thus while we can conclude that PL-Rank-2 has the best sample-efficiency of all the methods, the improvements over PL-Rank-1 and the placement policy gradient are most substantial when $N < 100$.

Overall, we see that the basic policy gradient and LambdaLoss are poor choices for optimization,
despite the fact that these are the methods we find in previous work~\citep{singh2019policy, bruch2020stochastic}.
The choice between PL-Rank-1 and the placement policy gradient does not seem to matter when only the number of epochs is considered.
PL-Rank-2 appears the safest choice because, in all tested cases, it either outperforms or has comparable performance to the other methods.

To conclude, we answer \ref{rq:sample} in the affirmative: PL-Rank-2 is the most sample-efficient method, although when $N > 100$ PL-Rank-1 and the placement policy gradient have comparable performance.

{\renewcommand{\arraystretch}{0.01}
\begin{figure*}[tb]
\centering
\begin{tabular}{r r r}
 \multicolumn{1}{c}{\small Yahoo! Webscope}
&
 \multicolumn{1}{c}{\small MSLR-WEB30k}
&
 \multicolumn{1}{c}{\small  Istella}
\\
\includegraphics[scale=0.38]{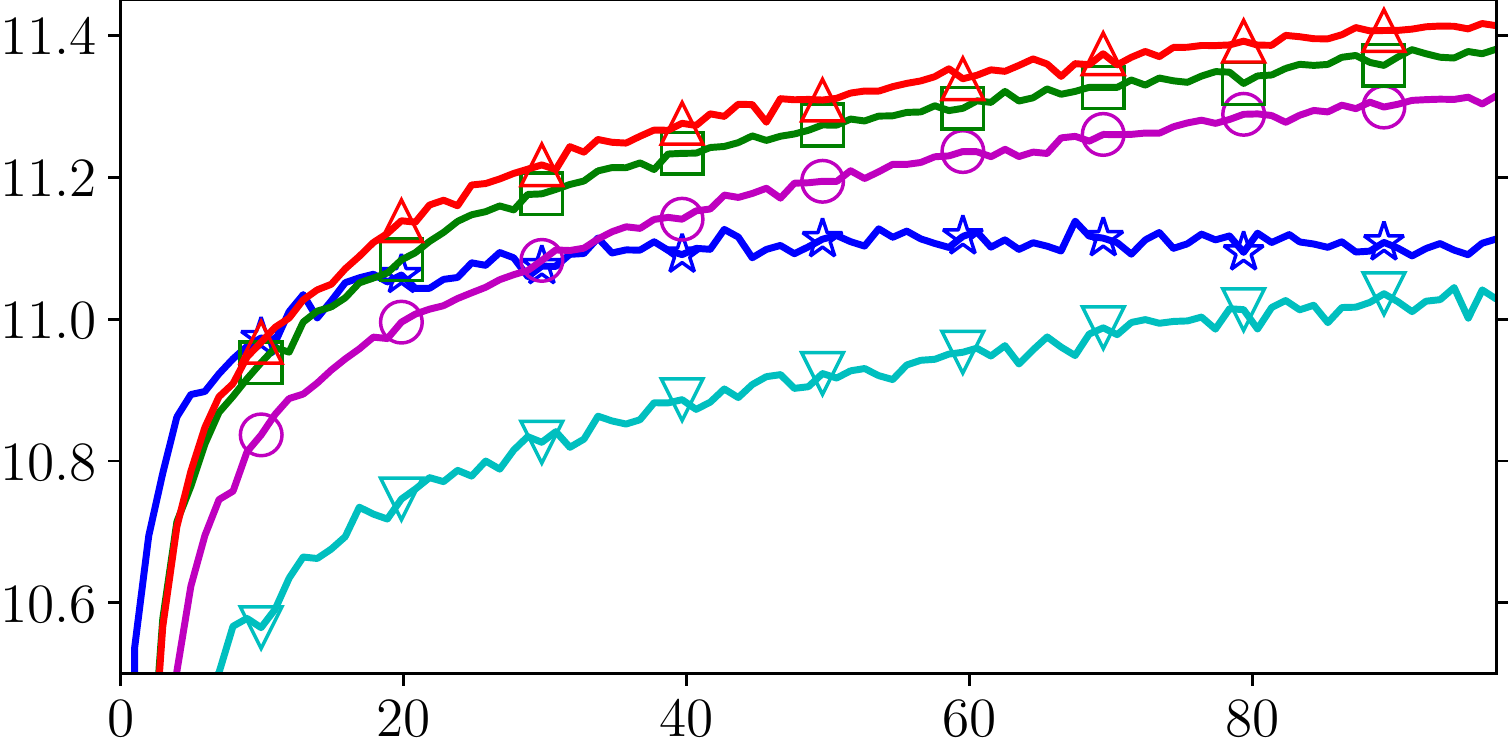} &
\includegraphics[scale=0.38]{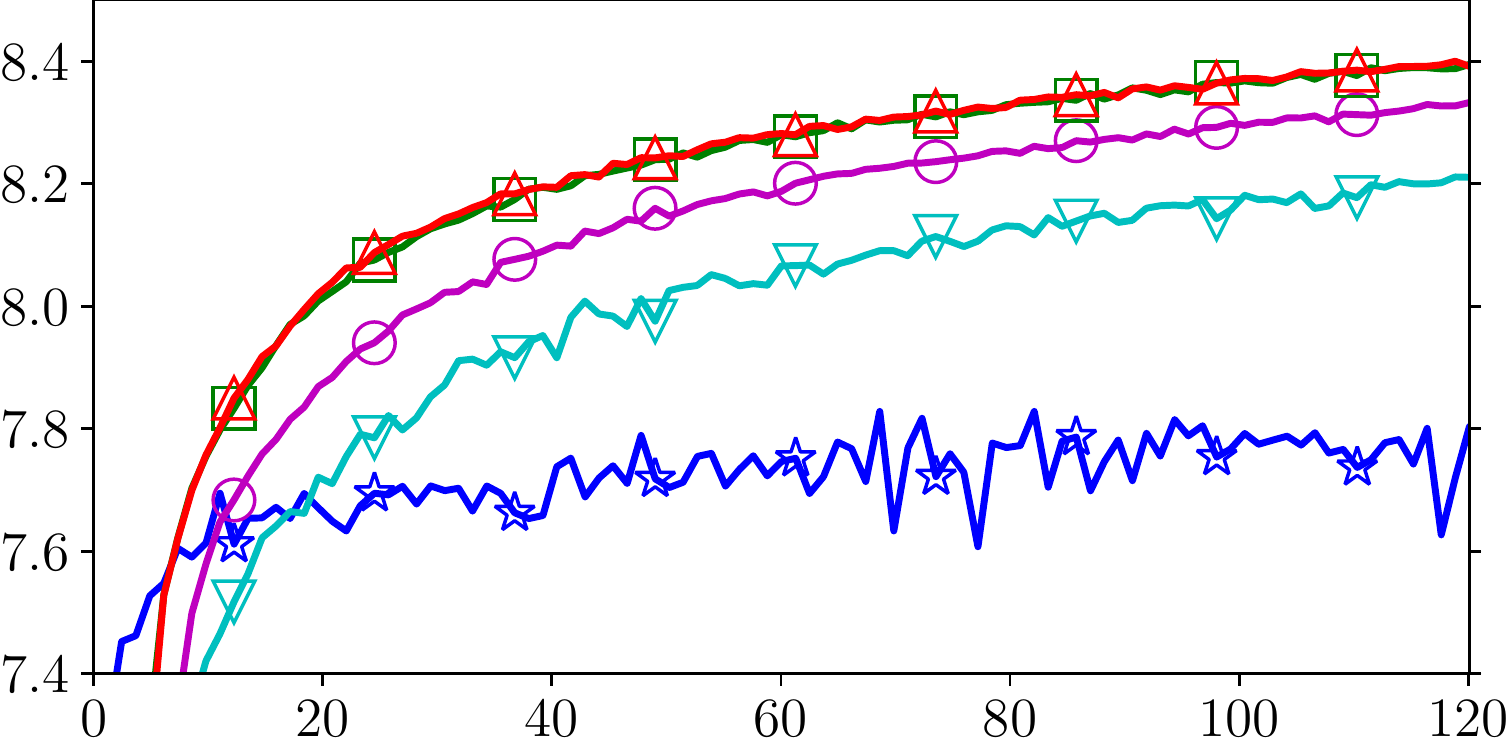} &
\includegraphics[scale=0.38]{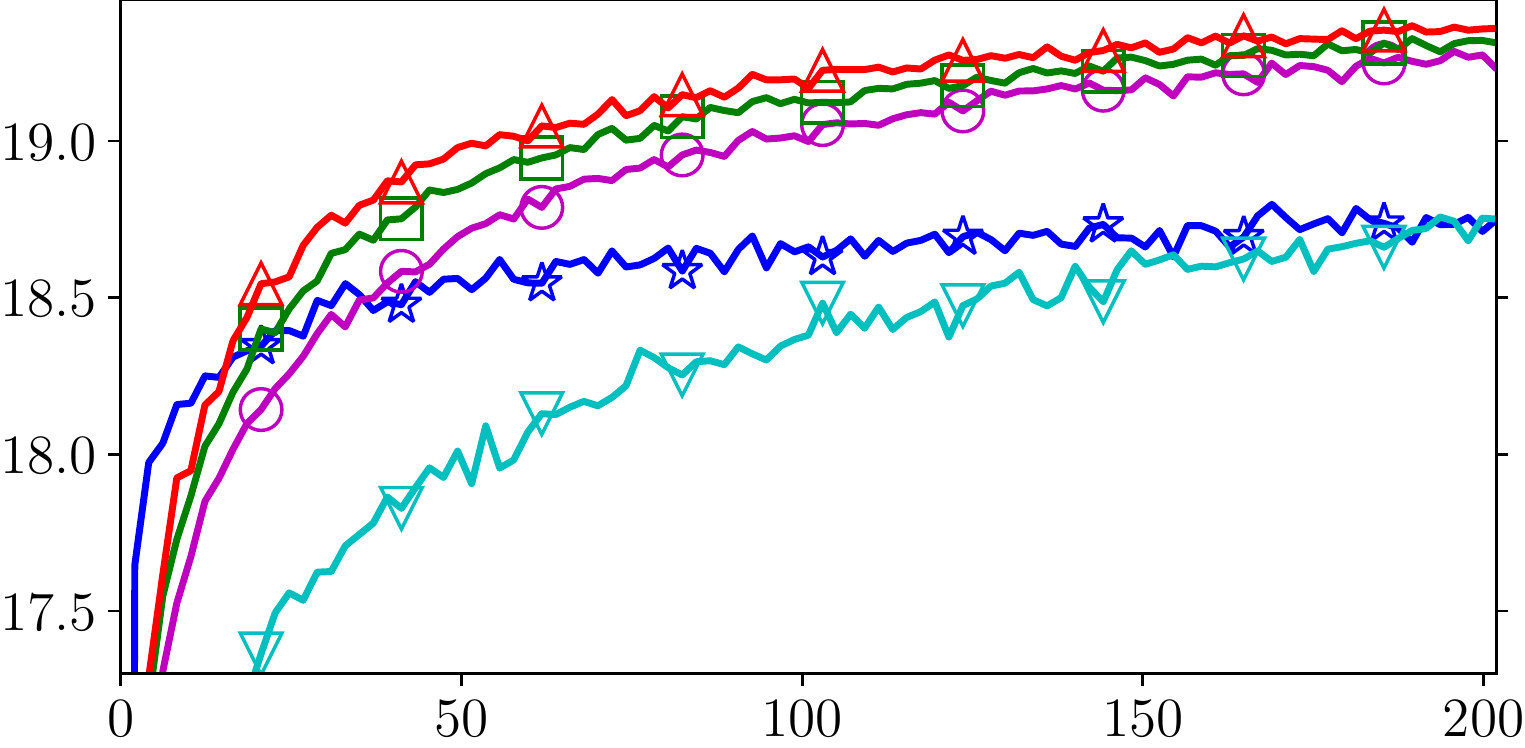} 
\\
\multicolumn{1}{c}{\small \hspace{0.5em} Minutes Trained}
& \multicolumn{1}{c}{\small \hspace{0.5em} Minutes Trained}
& \multicolumn{1}{c}{\small \hspace{0.5em} Minutes Trained}
\\
\multicolumn{3}{c}{
\includegraphics[scale=.45]{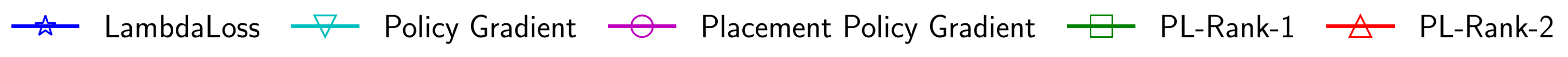}
} 
\end{tabular}
\caption{
Performance in DCG@5 of PL ranking models trained using different gradient estimation methods following a dynamically updated number of sampled rankings $N$ per estimation.
Results are the mean of 20 independent runs.
}
\label{fig:dynamic}
\vspace{-0.5\baselineskip}
\end{figure*}
}

\begin{table}
\caption{DCG@5 reached using different gradient estimation methods following a dynamically updated $N$ and being optimized for the same amount of time.
Results are the mean of 20 independent runs, the standard deviation is displayed in brackets, $\triangledown$ indicates the result is significantly worse ($p < 0.01$) than that of PL-Rank-2 on the same dataset.}
\label{tab:dcg}
\begin{tabular}{l c c c}
& Yahoo
& MSLR
& Istella
\\ \hline
Minutes Optimized
& 100
& 120
& 200
\\ \hline
LambdaLoss
& 11.11 {\tiny$^{\tiny \triangledown}$ ( 0.05)}
& 7.80 {\tiny$^{\tiny \triangledown}$ ( 0.09)}
& 18.75 {\tiny$^{\tiny \triangledown}$ ( 0.10)}
\\
Policy Gradient
& 11.03 {\tiny$^{\tiny \triangledown}$ ( 0.04)}
& 8.21 {\tiny$^{\tiny \triangledown}$ ( 0.07)}
& 18.75 {\tiny$^{\tiny \triangledown}$ ( 0.10)}
\\
Placement Policy Gradient
& 11.31 {\tiny$^{\tiny \triangledown}$ ( 0.02)}
& 8.33 {\tiny$^{\tiny \triangledown}$ ( 0.04)}
& 19.23 {\tiny$^{\tiny \triangledown}$ ( 0.06)}
\\
PL-Rank-1
& 11.38 {\tiny$^{\tiny \triangledown}$ ( 0.03)}
& \bf 8.39 {\tiny$^{\tiny -}$ ( 0.04)}
& 19.31 {\tiny$^{\tiny \triangledown}$ ( 0.05)}
\\
PL-Rank-2
& \bf 11.42 {\tiny$^{\tiny -}$ ( 0.02)}
& \bf 8.39 {\tiny$^{\tiny -}$ ( 0.03)}
& \bf 19.38 {\tiny$^{\tiny -}$ ( 0.05)}
\\
\hline
\end{tabular}
\end{table}

{\renewcommand{\arraystretch}{0.1}
\begin{figure}[tb]
\centering
\begin{tabular}{c r}
\small Yahoo! Webscope
\\
\includegraphics[scale=0.38]{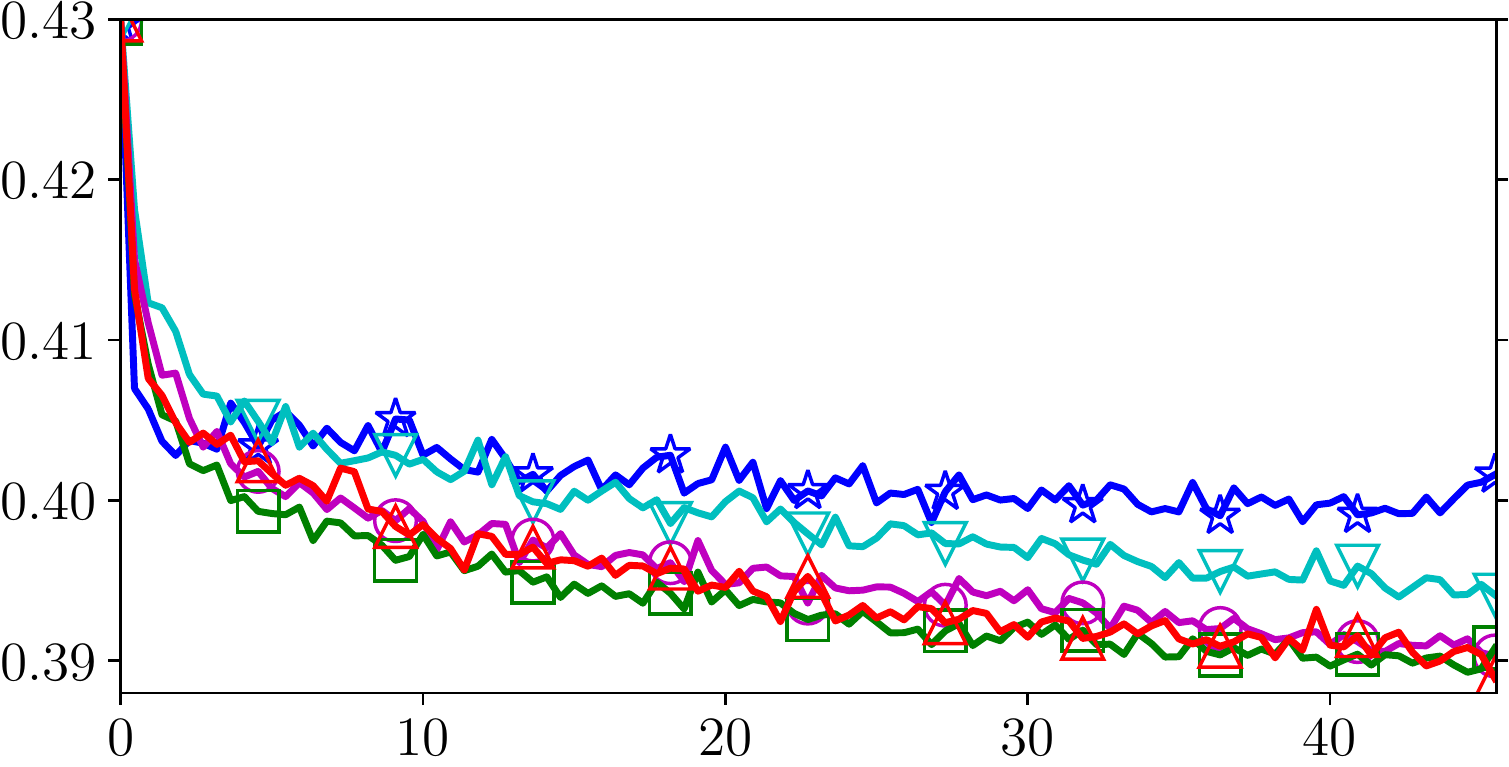} &
\hspace{-0.5em}
\includegraphics[scale=.42]{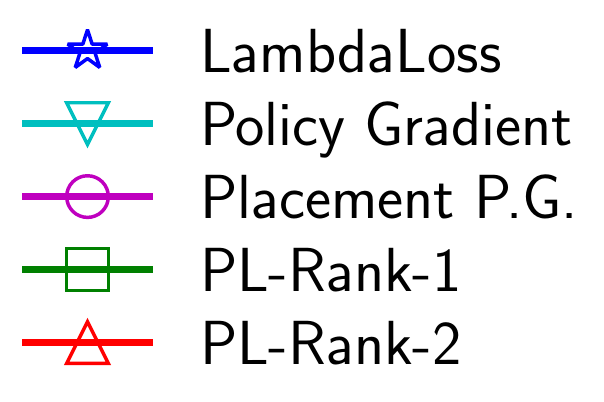}
\\
\small \hspace{0.5em} Minutes Trained
\\
\end{tabular}
\vspace{0.1\baselineskip}
\caption{
The mean disparity error of models trained using different gradient estimation methods following a dynamically updated number of sampled rankings $N$ per estimation.
Results are the mean of 20 independent runs.
}
\label{fig:fairness}
\vspace{-0.1\baselineskip}
\end{figure}
}

\subsection{Computational Costs and Time-Efficiency}
\label{sec:resultstime}
In order to answer \ref{rq:time}: \emph{whether PL-Rank requires less computational time to reach optimal performance}, we first consider Table~\ref{tab:time} which displays the average time taken to perform a single epoch per method for various $N$ values.

We see that in all cases PL-Rank-1 takes the least amount of time to compute, with PL-Rank-2 being the second fastest method.
There is little difference between the policy gradient methods but they are always much slower than the PL-Rank methods.
Depending on the dataset and $N$, the difference between PL-Rank and the policy gradients varies from around a minute to almost five minutes.
In Figure~\ref{fig:sample} we see that convergence requires at least 40 training epochs, thus differences in minutes per epoch can easily add up to reaching convergence over an hour earlier.

LambdaLoss is especially affected by the $N$ parameter, a likely explanation is that it is the only method that has to sample the complete ranking, whereas the other methods only need to sample the top-$k$ ranking (top-$5$ in this case).

By considering both the results from Table~\ref{tab:time} and Figure~\ref{fig:sample}, we can make three observations:
\begin{enumerate*}[label=(\roman*)]
\item when $N = 10$, decent but not optimal performance is reached;
\item $N = 100$ is enough to converge near optimal performance;
and \item performing an epoch with $N=10$ is considerably faster than with $N=100$.
\end{enumerate*}
Based on these observations, it seems reasonable to increase $N$ at every training step so that decent performance is reached very quickly but convergence is still optimal.
We found that $N = 10 +  90 \cdot \frac{\text{epoch}}{40}$ outperformed the static choices $N=10$ and $N=100$ in terms of learning speed while maintaining optimal convergence.

Figure~\ref{fig:dynamic} shows the performance of PL ranking models trained using this dynamic $N$ strategy over training time in minutes.
Again we see that LambdaLoss converges fast but at suboptimal performance.
Similarly, there is clearly a large difference visible between the basic policy gradient and the placement policy gradient.
However, on all datasets, we see an improvement of PL-Rank-1 over the placement policy gradient which is very large on Yahoo and MSLR but smaller on Istella.
This improvement can be attributed to the reduced computational costs of PL-Rank-1, as a result, it is capable of completing more epochs in the same amount of time and can therefore reach a higher performance in less computational time.
Finally, compared to PL-Rank-1, PL-Rank-2 has an even higher performance on the Yahoo and Istella datasets but not on MSLR.
It appears that its increased sample-efficiency helps PL-Rank-2 initially, when $N$ is low, except on the MSLR dataset where Figure~\ref{fig:sample} also shows us that the difference in sample-efficiency is very limited.
To better verify that PL-Rank-2 is the best choice, we performed a two-sided student t-test on the performance differences, the results are displayed in Table~\ref{tab:dcg}.
We see that Pl-Rank-2 is significantly better compared to the other methods in all tested cases, with the single exception of PL-Rank-1 on the MSLR dataset.

To conclude, we answer \ref{rq:time} in the affirmative: the PL-Rank methods achieves significantly higher performance with less computational time required than LambdaLoss or the policy gradients.
In particular, PL-Rank-2 is the most time-efficient method across all datasets.

\subsection{Optimizing a Ranking Fairness Metric}
Finally, we address \ref{rq:fair}: \emph{whether PL-Rank is effective at optimizing ranking fairness}.
Figure~\ref{fig:fairness} displays the mean disparity error for models optimized with the different gradient estimation methods, with the same increasing $N$ strategy applied as in Section~\ref{sec:resultstime}.
While all methods decrease the disparity, the PL-Rank methods and the placement policy gradient are considerably more efficient than LambdaLoss and the basic policy gradient.
Unlike with the optimization for relevance, there appears only a very small improvement of the PL-Rank methods over the placement policy gradient.
Therefore, we answer \ref{rq:fair} in the affirmative: PL-Rank can effectively optimize ranking fairness, where we note that the placement policy gradient has comparable time-efficiency.

\section{Conclusion}
\label{sec:conclusion}

In this paper, we tackled the optimization of \ac{PL}-ranking models for both relevance and fairness ranking metrics.
While previous work has found PL-ranking models effective for various ranking tasks, their optimization can involve large computational costs.
To alleviate these costs, we introduced three new estimators for efficiently estimating the gradient of a ranking metric w.r.t.\ a PL ranking model: the placement policy gradient and two PL-Rank methods.
The latter two can be computed using the PL-Rank algorithm.
To the best of our knowledge, PL-Rank is the first algorithm designed specifically for efficiently optimizing PL ranking models w.r.t.\ ranking metrics.
Our experimental results indicate that our novel methods considerably reduce the computational time required to reach optimal performance compared to existing methods.
In particular, the PL-Rank-2 method has the best sample-effiency and was found to reach significantly higher performance when ran for the same amount of time as other methods.
Compared to the popular basic policy gradient, PL-Rank-2 converges several hours earlier, thus immensely alleviating the computational costs of optimization.

With the introduction of PL-Rank, we hope that the usage of stochastic ranking models is made more attractive in real-world scenarios. %
Finally, we think PL-Rank is also an important theoretical contribution to the \ac{LTR} field, as it proves that PL ranking models can be optimized with computational efficiency, without relying on heuristic methods.

\subsection*{Code and data}
To facilitate reproducibility, this work only made use of publicly available data and our experimental implementation is publicly available at \url{https://github.com/HarrieO/2021-SIGIR-plackett-luce}.

\clearpage

\balance
\bibliographystyle{ACM-Reference-Format}
\bibliography{references}

\end{document}